\documentclass[%
preprint,
 amsmath,amssymb,
 aps,
]{revtex4-2}

\usepackage{graphicx}% Include figure files
\usepackage{dcolumn}% Align table columns on decimal point
\usepackage{bm}% bold math
\usepackage{hyperref}% add hypertext capabilities

\begin{document}

\title{Boundless metamaterial experimentation: physical realization of a virtual periodic boundary condition}

\author{Henrik R. Thomsen}
\email{henrik.thomsen@ibk.baug.ethz.ch}

\author{Bao Zhao}%

\author{Andrea Colombi}
\affiliation{
  ETH Zurich, Department of Civil, Environmental and Geomatic Engineering,
  Stefano-Franscini-Platz 5, 8093 Zurich, Switzerland
}

\date{\today}

\begin{abstract}
We experimentally implement a virtual geometric periodicity in an elastic metamaterial. First, unwanted boundary reflections at the domain ends are cancelled through the iterative injection of the polarity reversed, reflected wavefield. The resulting boundless experimental state allows for a much better analysis of the metamaterials influence on the propagating wavefield. Subsequently, the propagating wavefield exiting on one end of the structure is reintroduced at the opposite end, creating a virtual geometric periodicity. We find that the experimentally observed band gap converges to the analytical solution through the introduction of the virtual periodicity. The established workflow introduces a novel approach to the experimental investigation and validation of metamaterial prototypes in the presence of strongly dispersive wave propagation and internal scattering. The fully data driven, ad-hoc treatment of boundary conditions in metamaterial experimentation with arbitrary mechanical properties enables reflection suppression, virtual periodicity, and the introduction of more general fictitious boundaries. 
\end{abstract}
                          %display desired
\maketitle

\section{Introduction}

Engineered metamaterials have long since proven their capabilities to exhibit extraordinary material properties \cite{Bertoldi2010,Zheng2014,Wang2021}, as well as manipulate and attenuate wave propagation \cite{Colombi2016,Matlack2016} using band gaps. Such prohibited frequency ranges can be architected into materials via either Bragg scattering phononic crystals \cite{sigalas1992elastic} or locally resonant metamaterials \cite{liu2000locally}. The later enables sub-wavelenghth bandgaps \cite{craster2012acoustic} and has recently been leveraged for broadband energy harvesting \cite{DePonti2020experimental,DePonti2020graded,DePonti2021enhanced,Bao2022}. Numerical simulations drive the development of metamaterials, the ever increasing available computational power making it feasible to rapidly model different designs. Nevertheless, experimental investigation and validation of prototype structures remains essential. Whereas in numerical studies, effects such as low absorbing boundary layers or local damping \cite{Komatitsch2007,RAJAGOPAL201230} can be introduced to simulate free space wave propagation, laboratory experiments are often plagued by modal responses of the system caused by, e.g., free or clamped boundaries. Additionally, wave propagation within the structure is difficult to interpret due to boundary reflections. Thus, it is often challenging to accurately distinguish phenomena caused by the metamaterial under study from those of the whole system. Conventional experimental solutions to these pitfalls are passive boundary reflection damping, such as acoustic black holes \cite{Mironov1988,Oboy2010,Georgiev2011} or graded impedance interfaces \cite{Vemula1996}. However, these practices result in modifications of the host structure which in turn can lead to deviations from the desired system response. 

A promising alternative is elastic “immersive wave experimentation” (IWE) \cite{Thomsen2019}, used to actively remove the unwanted boundary reflections at the domain ends of a solid target, such as an elastic metamaterial waveguide. IWE is an experimental methodology aimed at overcoming laboratory- and sample-size related limitations that affect conventional wave propagation experimentation. This is achieved by linking wave propagation between a physical experimental domain and a desired virtual domain through modification of the physical boundary conditions encountered in the laboratory \cite{Vasmel2016b}. The method can be used in a wide range of applications, such as cloaking and holography \cite{Boersing2019}, as well as virtual extension \cite{Thomsen2019} and modification \cite{Becker2020} of the finite physical domain encountered in the laboratory. It has also been proposed to utilize acoustic IWE in implementing arbitrary experimental periodic boundary conditions by cancelling reflections on one end of a structure and re-introducing the propagating wavefield on the opposite side \cite{vanManen2019}. If successfully employed this would allow to study the effect of, e.g., phononic crystals or locally resonant metamaterials without the need to physically realize large structures in the laboratory but instead only a few unit cells. Integral to IWE is the correct identification of the first-order incident event at the domain boundaries, i.e., wave energy which did not interact with said boundaries before. This event is then used to construct a secondary force signature which when applied at, e.g., the free ends of an aluminum beam, leads to the cancellation of the reflected energy, creating a transparent boundary which can then be fully controlled and linked to an arbitrary virtual domain, if desired.  

Our work demonstrates the first use-case of elastic IWE and significantly expands on the method in two key aspects: (1) the frequency range of interest is $<$1 kHz, rather than tens of kHz and (2) we focus on flexural (Lamb) waves propagating along a graded beam characterized by strong dispersion and internal scattering. Such wave propagation is often encountered in metamaterial research and isolating the first-order reflected wavefield cannot be easily achieved due to the strong dispersion and scattering. In this context, the cancellation of reflections originating at the ends of the structure, allows to clearly analyze the modulation of the wave when passing through, e.g., the graded area of a prototyped metamaterial. 

In the following, we will first introduce our proposed workflow to implement elastic ``immersive boundary control'' in the experimental domain. We schematically showcase how to effectively cancel boundary reflections in three cases often encountered in phyiscal experiments, with a focus on metamaterials: (1) the base case, single side IWE, where a clear identification of the incident/reflected wavefield at the structures boundaries is possible, (2) iterative, single side IWE, where the clear identification of the incident and/or reflected wavefield is not possible due to e.g., the presence of strong dispersion, and (3) double sided IWE, with internal scattering caused by strong impedance contrasts due to, e.g., the design of the metamaterial, requiring, again, the utilization of an iterative procedure. The proposed workflow is then used to successfully remove unwanted boundary reflections from a metamaterial designed for energy harvesting purposes \cite{Bao2022} and to create a virtual geometric periodicity in the elastic experiment.

\section{Iterative procedure to implement immersive boundary conditions}

Elastic IWE states that applying the incident traction as an auxiliary force on the free surface of an experimental domain, such as the face of a rock or the ends of an aluminum beam, results in a cancellation of the wavefield reflected at said boundary, accounting for any mode conversions. \citet{Thomsen2019} showed that this effect can be implemented experimentally by applying the polarity flipped, first-order, reflected wavefield at the free surface ends of the physical domain under investigation. Allowing for the cancellation of both reflected longitudinal and flexural wave modes. A first order reflection describes an event that did not interact with the domain ends before. Different to acoustic IWE, as presented by e.g., \citet{Becker2020} and \citet{Boersing2019}, elastic IWE is nearly impossible to implement in real-time because of the high wave propagation velocities. This leads to an unfeasible small lag time to predict the incident wavefield before it reaches the boundary. Therefore, the desired immersive experimental state is built up iteratively. Key is the correct identification of the incident and reflected wavefield. Possible methods to do so are e.g., wavefield injection \cite{Thomsen2021}, redatuming \cite{Sonneland2005} and f-k domain filtering \cite{Hardage1985}. The latter is used in our proposed workflow. 

\begin{figure}[ht!]
\begin{center}
\includegraphics[]{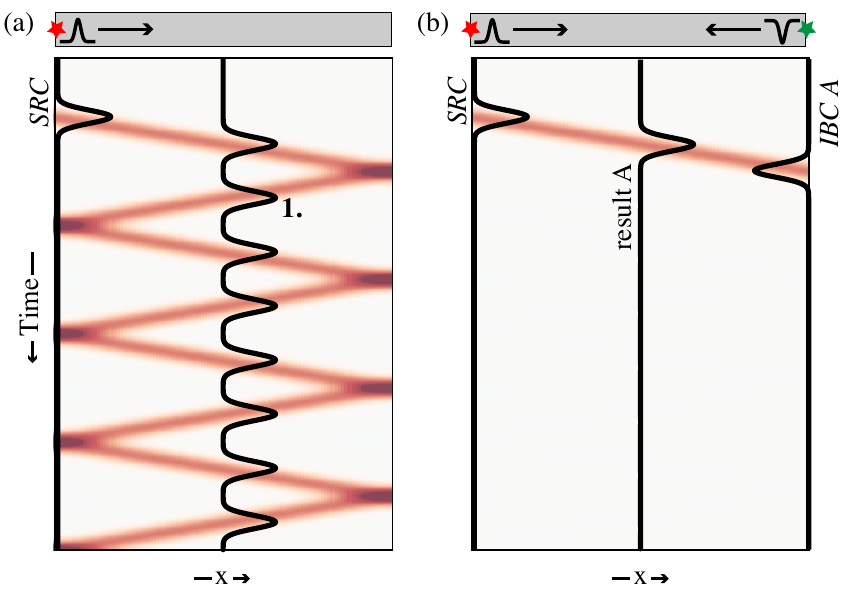}
\caption{Conceptual representation of time-distance plots for Case 1: single sided and clear incident/reflected wave identification: (a) displays the initial experimental state and (b) the result when cancelling the reflected events by injecting the polarity reversed first order reflected event at the right end. The applied IBC is indicated by a black trace on the right end of the plot.\label{fig:fig01}}
\end{center}
\end{figure}

\subsection{Case 1: single side IBC and clear incident/reflected wave identification}

The most straightforward procedure to implement immersive boundary control (IBC) is through injection of the polarity reversed, first order, reflected wave measured at the ends of the experimental domain. Figure \ref{fig:fig01}(a) schematically shows a 1D example of a longitudinal wave propagating back and forth between two free ends over time. The wavefield is induced on the left side by the source (\textit{SRC}). Identifying and isolating the first event reflected at the right end (indicated by the 1.) results in the design of the IBC: the polarity reversed version of said event. When injected on the right, as indicated by \textit{IBC A} in Fig. \ref{fig:fig01}(b), the first reflection occurring is effectively cancelled and any higher order reflections cease to exist. This procedure is used if the first order reflected event can be clearly identified. 

\begin{figure}[ht!]
\begin{center}
\includegraphics[]{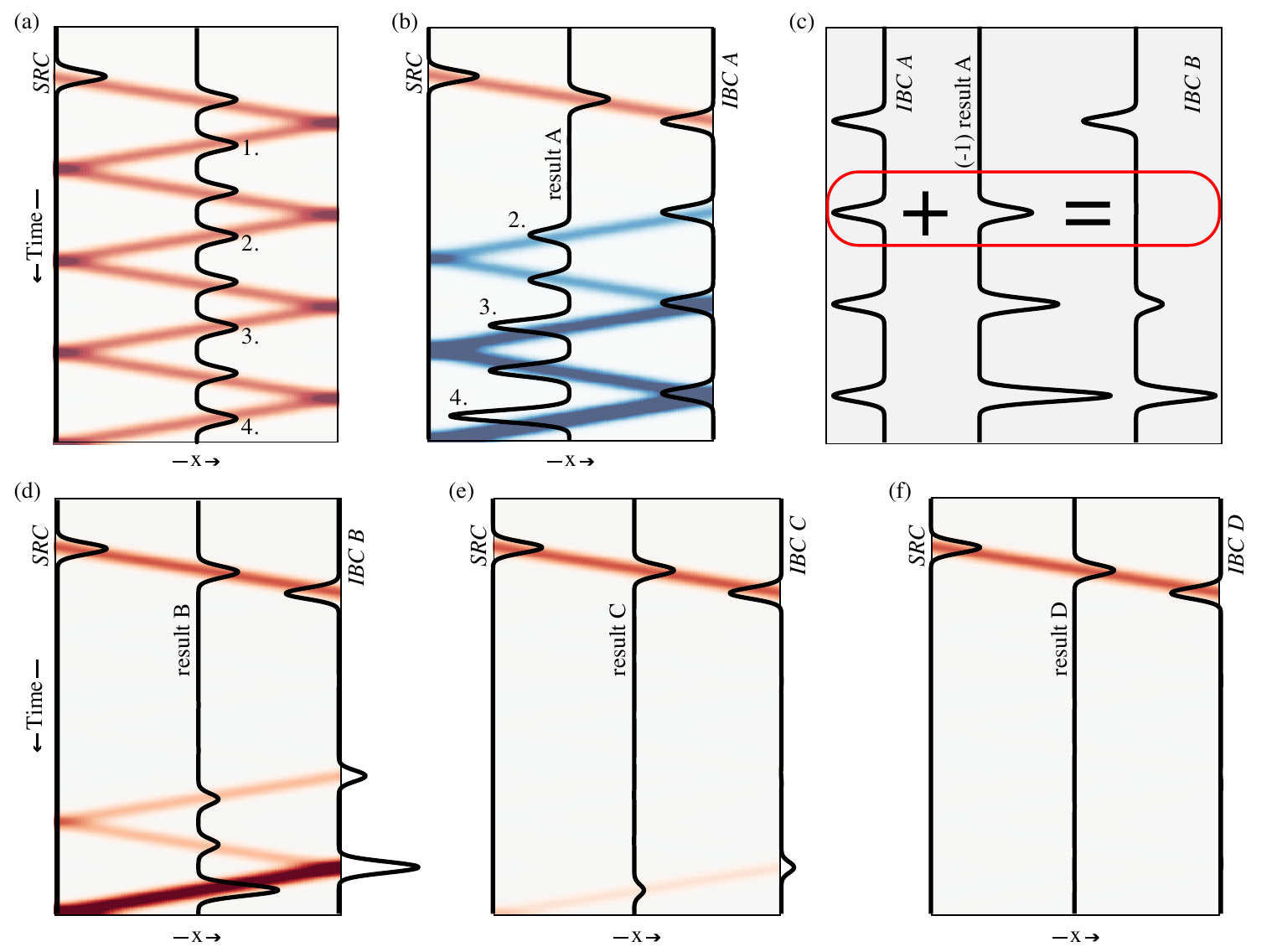}
\caption{Conceptual representation of time-distance plots for Case 2: single sided and iterative procedure to remove boundary reflections: (a) displays the initial experimental state, (b\&d-f) the evolution of the IBC applied to the right end, as indicated by the black trace, as well as the resulting wave propagation and (c) how the IBC is iteratively modified by adding the currently applied IBC and the resulting polarity reversed reflected wavefield. \label{fig:fig02}}
\end{center}
\end{figure}

\begin{figure}[ht]
\begin{center}
\includegraphics[]{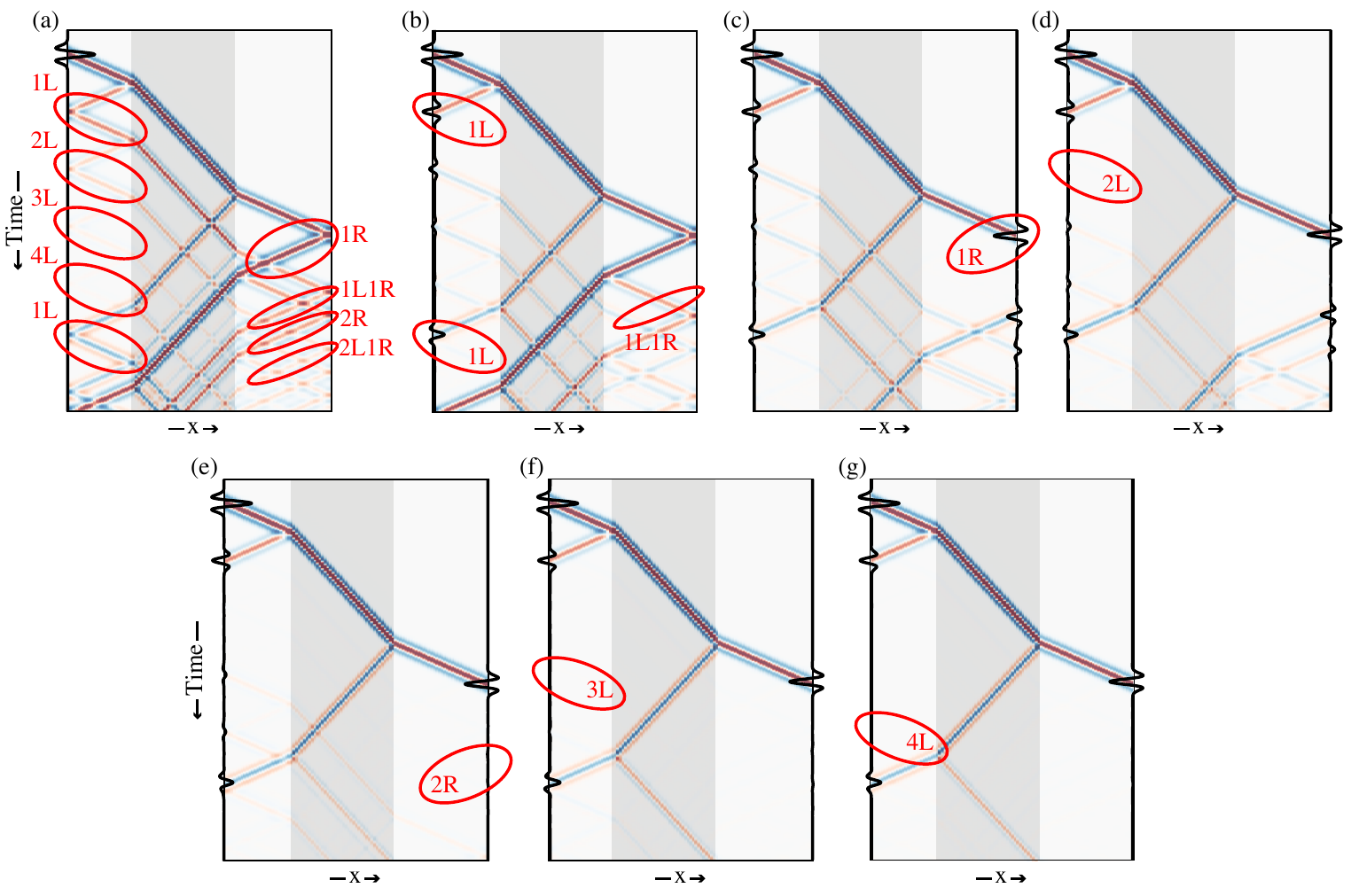}
\caption{Conceptual illustration of time-distance plots for Case 3: double sided IBC with internal scattering due to an impedance contrast indicated in gray: (a) displays the initial experimental state, (b-f) the evolution of the IBC applied to the right and left ends, as well as the resulting wave propagation, (g) the final experimental state after the left and right IBC converge to only applying the polarity reversed first order events and resulting primary wavefield propagation. \label{fig:fig03}}
\end{center}
\end{figure}

\subsection{Case 2: single sided IBC and iterative procedure}

Unfortunately, a clear identification of just the required first order reflected event is not always possible, especially when dealing with strongly dispersive wave propagation or internal reflection. In such cases, the desired boundary control can still be achieved through the use of the iterative procedure proposed herein.

Starting with the initial experimental state displayed in Fig. \ref{fig:fig02}(a), rather than injecting just the polarity reversed version of the first reflected event, the \textit{full} reflected wavefield, labeled 1.-4., i.e. the wavefield propagating to the left, is injected as indicated by \textit{IBC A} on the right end of Fig. \ref{fig:fig02}(b). As a result, the first order reflected event is removed as can be seen in the same Figure. This results in all higher order events to cease existing as in Case 1. However, by applying the polarity reversed \textit{full} reflected wavefield new energy is actively injected into the system. Thus, the second order event can now be found propagating at opposite polarity to the initial experimental state. Additionally, the wavefield is amplified for every higher interaction with the right end due to the constructive interference of the propagating wavefield with the continued injection of higher order reflected events as part of the initial \textit{IBC A}. These unwanted events are eventually removed by repeating the experiment and using the consecutive experimental states to iteratively modify the IBC. The final IBC only applies the first order reflected wavefield without introducing unwanted wavefield propagation. 

This is achieved by isolating the new wavefield reflected at the right end as indicated by 2.- 4. in Fig. \ref{fig:fig02}(b), reversing it in polarity and adding it to the previously applied \textit{IBC A}. The resulting \textit{IBC B} is shown in Fig. \ref{fig:fig02}(c). Here, the second order event is removed from the trace of \textit{IBC B}. Thus, one higher order event less is reintroduced to the experimental state during the application of \textit{IBC B}. Figure \ref{fig:fig02}(d) then shows how \textit{IBC B}, when applied, removes both the first and second order reflections, yet still introduces the third and amplified higher order events. Finally, by repeating the process an additional two times, \textit{IBC D} in Fig.~\ref{fig:fig02}(f) matches \textit{IBC A} from Fig. \ref{fig:fig01}(b) in the time range shown. 

In summary, the iterative procedure removes the unwanted higher order reflected events step-by-step. The IBC, initially containing the \textit{full} reflected wavefield, converges to the first order reflected wavefield.

\subsection{Case 3: double sided IBC with internal scattering present}

The iterative method to construct the IBC described above can also be used when reflections occur not only at the ends but also within the structure as schematically displayed by Fig. \ref{fig:fig03}(a). In this example an internal impedance contrast, indicated by the gray shaded area, causes internal scattering. Such cases are commonly encountered when investigating metamaterials where, e.g., resonators introduce frequency dependent impedance contrasts in the medium. The internal impedance contrast results in two types of undesired boundary reflections: those transmitted through the structure and reflected at the ends and those radiated internally and then interacting with the end boundaries.

Therefore, the construction of the IBC requires a double sided, iterative procedure to construct the desired boundless experimental state where only primary wave propagation remains. In Fig. \ref{fig:fig03}(a) one can observe first order reflections on both the left and right ends, as indicated by \textit{1L} and \textit{1R} respectively, as well as higher order events (\textit{2L-4L}, \textit{2R} and \textit{1L1R-2L1R}). In a first step the complete wavefield reflected at the left end is reversed in polarity and injected together with the source wavefield. The result, displayed in Fig. \ref{fig:fig03}(b), shows how both first order reflections (\textit{1L}) are removed on the left side. The higher order events are again amplified as previously described in Case 2. Additionally, the reflection on the right side (event \textit{1L1R}), originating from event \textit{1L}, is also removed in Fig. \ref{fig:fig03}(b). Next, the full reflected wavefield acquired on the right side of the new experimental state, displayed in Fig. \ref{fig:fig03}(b), is reversed in polarity and injected at the right end. As a result, the reflection of event \textit{1R} is cancelled and event \textit{2R} is flipped in polarity, as can be observed in Fig.~\ref{fig:fig03}(c). Next, the IBC on the left end is updated with the new experimental state. Resulting in the removal of event \textit{2L}, as shown in Fig.~\ref{fig:fig03}(d). This iterative process of applying and then constructing the IBC wavefield on opposite sides is repeated several more times (Fig. \ref{fig:fig03} (e-g)). Until the left and right IBC wavefields, displayed as black time-traces at the ends of Fig.~\ref{fig:fig03} (g), only contain the first order reflected events when compared to the initial IBC shown in Fig.~\ref{fig:fig03}(c). The final boundless experimental state when applying the dual sided IBC is shown in Fig.~\ref{fig:fig03}(g). All undesired reflections at the boundaries have been removed and only the propagation of the primary scattered wavefield remains.

Note that the iterative procedure described cancels the unwanted boundary reflections up to a cutoff time defined by the iteration step, i.e., the order of reflected events currently removed. Wave propagation after this time is amplified due to the constructive interference of the injected IBC wavefield and propagating waves. Thus, increased damping in the system under investigation can be advantageous as it would reduce the required iteration steps to achieve the desired experimental state. 

\begin{figure}[ht]
\begin{center}
\includegraphics[]{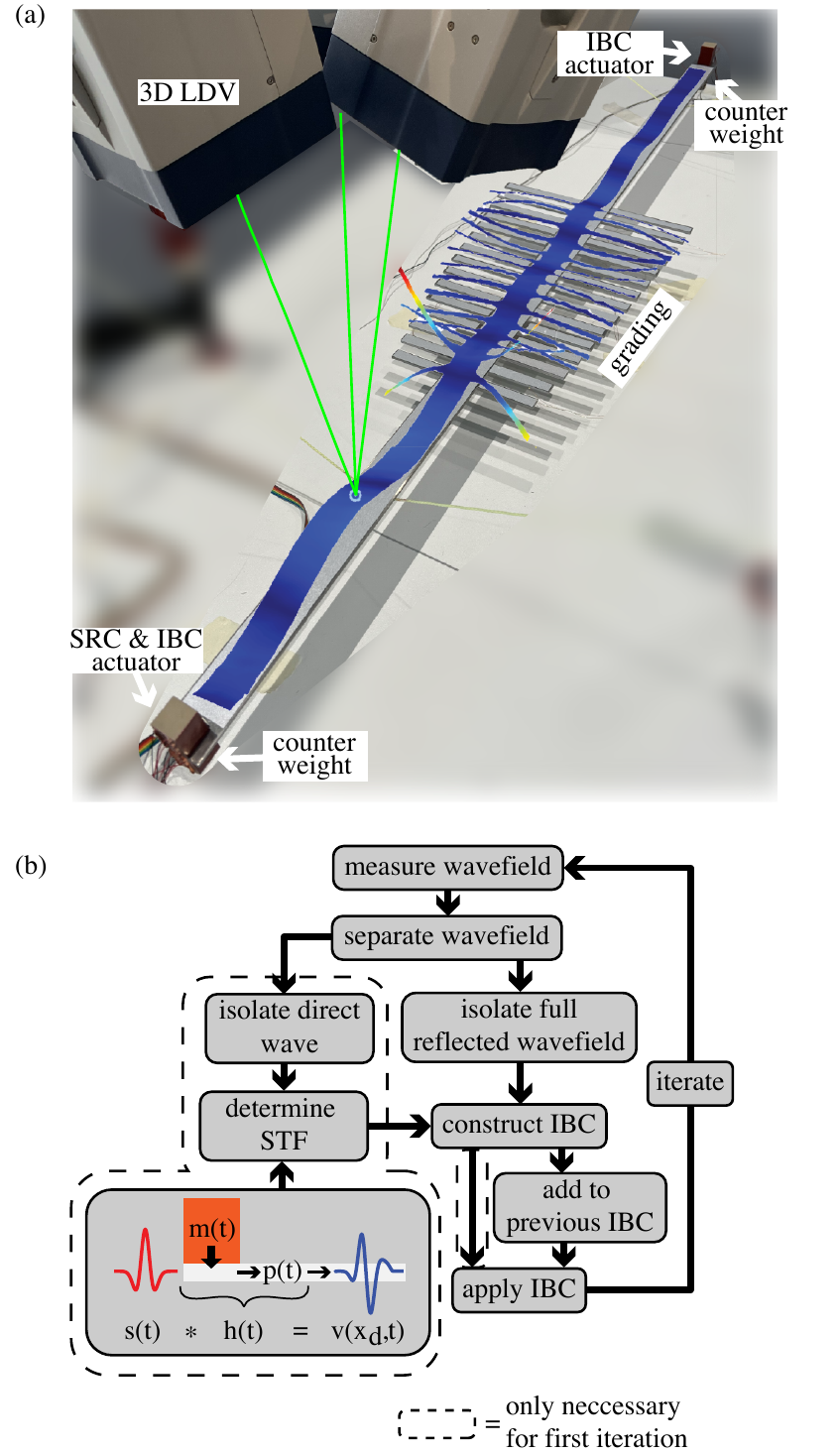}
\caption{(a) Experimental set-up. A piezoelectric actuator and counter weight are attached to both ends of the beam. The 3D LDV is used to measure the out-of-plane wavefield propagation. Overlain is a snapshot of the recordings along the full structure at 12.88 ms. (b) Workflow to iteratively construct and apply the IBC. The dashed lines indicate steps only required for the first iteration. \label{fig:fig04}}
\end{center}
\end{figure}

\section{Experimentally implemented virtual periodic boundary}

We will now experimentally demonstrate the implementation of a virtual periodic boundary using IWE to aid in the investigation of a graded metamaterial. The prototype structure under investigation is shown in Fig. \ref{fig:fig04}(a). It is made up of a 2 millimeters thin, one-meter-long, host beam endowed with ''wings`` of increasing length arranged in the central region spanning 30 centimeters (from $0.3$ - $0.6$ m). The wings form a graded metasurface which acts as a spatial frequency filter on the propagating wave whilst simultaneously slowing down the group velocity of the traveling wave. This metamaterial was designed by \citet{Bao2022} for broadband and high-capability energy harvesting at low frequency ambient noise vibrations. First, we constructed and applied an IBC on the right and left ends of the host beam to cancel the reflections and create transparent boundaries. Following this, the wavefield incident on the right end was reintroduced to the structure on the left end to create a virtual geometric periodicity in the metamaterial.

\subsection{The metamaterial under investigation}

\begin{figure}[ht]
\begin{center}
\includegraphics[]{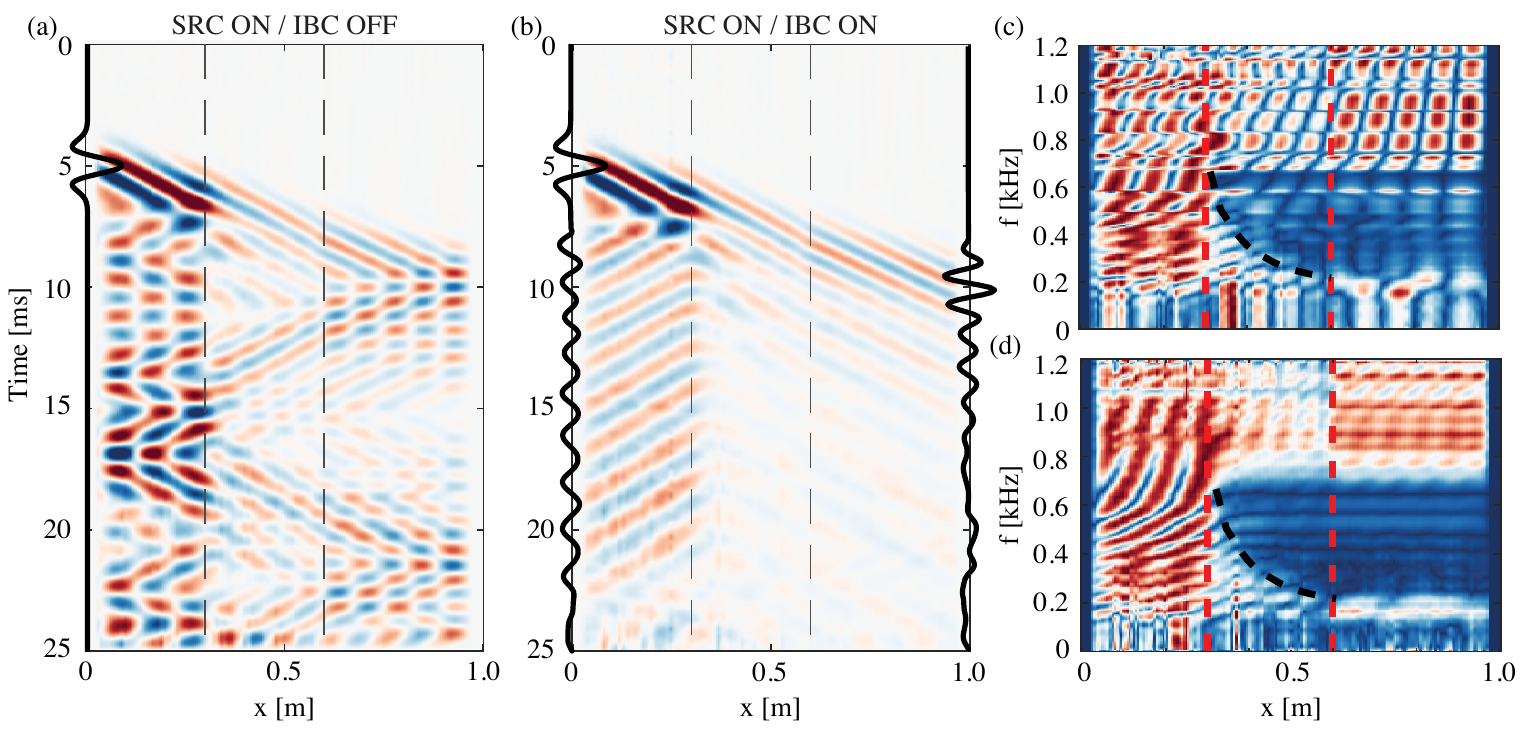}
\caption{Time-distance representation of the out-of-plane velocity measured along the beam. The vertical dashed lines indicate the graded area. (a) Source ON (illustrated by the black line on the left) and IBC OFF, (b) Source ON and IBC ON (illustrated by the black line on the left and right). (c-d) Spatial frequency distribution, normalized for each frequency, along the structure with the IBC OFF and ON, respectively. The black dashed lines track the band gap. Animations of the wave propagation along the structure, scanned by the 3D LDV, for both (a) and (b) can be found in the SUPPLEMENTARY MATERIAL (Movie 00 and 01). \label{fig:fig05}}
\end{center}
\end{figure}

A flexural plane wave was induced on the left end by a piezoelectric actuator (P. 153.05 \cite{Pica}). The same type of actuator was also attached to the beam on the opposite end. Both actuators act as IBC sources to control their respective boundaries. Additionally, two counterweights of $~$44.3 g were attached to the respective ends of the beam. Together with the actuator, they created a symmetric boundary condition. Testing showed that when only the piezoelectric actuator was attached to the ends of the beam, undesired mode conversions occurred during wave reflection. Adding the actuators modifies the boundary conditions at the ends of the beam. However, contrary to conventional reflection mitigation methods, we retain full control of the boundary response through the application of the IBC. The out-of-plane velocity was measured in the time domain along the beam using a robotized 3D LDV \cite{Polytec} as illustrated in Fig. \ref{fig:fig04}(a). The data was recorded for 100 ms with a sampling frequency of $f_s =$100 kHz and averaged five times at each of the 101 measurement points spaced 1 cm apart along the beam. A 500 Hz (central frequency) Ricker wavelet, with peak amplitude of 100 V, was driven to the out-of-plane component of the actuator placed on the beam's left end. Figure \ref{fig:fig05}(a) depicts a time-distance representation of the resulting out-of-plane flexural wave ($A_0$ Lamb mode) travelling along the beam for the initial 25 ms. The strong dispersive nature of the propagating flexural wave, as well as the strong internal scattering occurring due to the large impedance contrast imposed by the graded area, are apparent. 

The effect of a metamaterial on the propagating wave is commonly investigated through analysing the normalized spatial frequency distribution along the structure as depicted in Fig. \ref{fig:fig05}(c). The length of the wings attached to the host beam evolves from short to long. Consequently, higher frequencies, are filtered out earlier along the graded area than the lower frequencies. Thus, the frequency content of the wavefield incident on the graded area is modified through interaction with it. A phenomenon referred to as rainbow trapping \cite{Tsakmakidis2007,Colombi2016,DePonti2021}. Overall the metamaterial induces a band gap from 200 to 680 Hz. Noticeable is a peak in the spatial frequency content at 585 Hz, protruding into the band gap. As we will show later, this mode shape is a result of the boundary reflections occurring at the ends of the host beam.

\begin{figure}[ht]
\begin{center}
\includegraphics[]{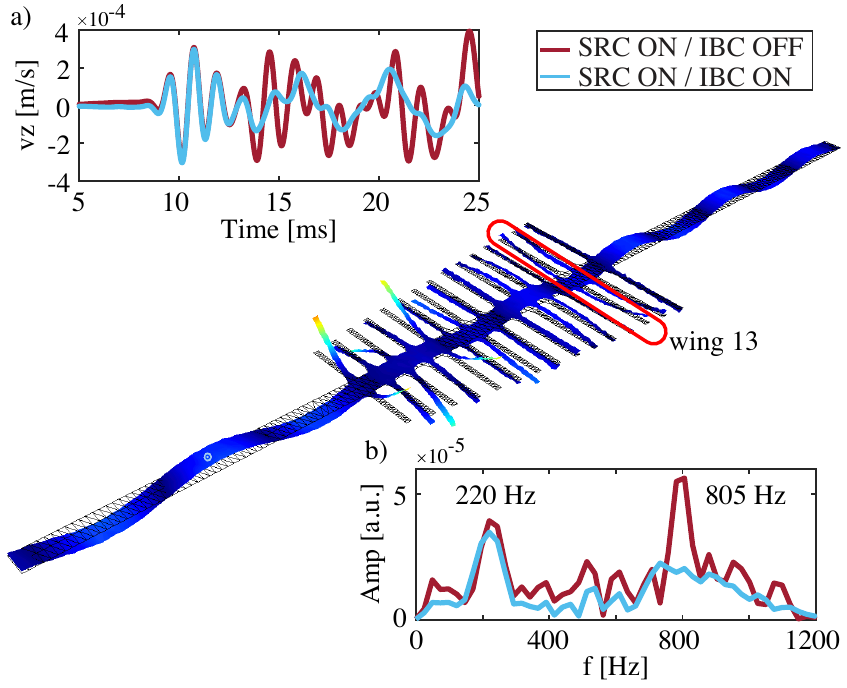}
\caption{Snapshot of the wave propagation recorded via the 3D LDV at 19.50 ms. Shown are the (a) time series and (b) frequency spectra of the out-of-plane velocity recorded at the tip of wing 13. The red (dark) lines indicate the experimental state with boundary reflections present (SRC ON / IBC OFF) and the blue (light) lines those with cancelled boundary reflections (SRC ON / IBC ON). \label{fig:fig06}}
\end{center}
\end{figure}

A clear understanding of the frequencies at which the individual wings comprising the graded metastructure resonate at is also crucial during the experimental investigation of the prototype. Such insights are important when optimizing the grading profile and resulting band gap. As an example the red (dark) lines in Fig. \ref{fig:fig06}(a\&b) depict both the out of plane velocity and frequency spectrum measured at the tip of wing 13. Strong peaks in the frequency response can be observed at both 220 and 805 Hz, showing how the reflections at the ends of the structure can make a clear identification of the resonance frequency difficult. 

\subsection{Step 1: reflection removal}
We used Case 3 described above to remove the undesired boundary reflections, i.e. the IBC was constructed iteratively on both ends. First, the full wavefield reflected at the left and right end had to be determined and applied as the IBC. Hence, the out-of-plane component ($v(x,t)$) measured along the beam was separated into its right ($r(x,t)$) and leftward ($l(x,t)$) propagating components. In the f-k domain representation of the data, the right- and leftward propagating wavefield map into opposite quadrants and can be separated using a masking filter \cite{Hardage1985}. However, before applying the full reflected wavefield at the respective end of the beam, and iteratively realizing the IBC, an additional step is required when working with laboratory data. As the IBC is applied via the two actuators at the ends, we must first determine which voltage signal $s(t)$ to apply to the actuators to induce the desired wavefield propagation $v(x_d,t)$ at any point $x_d$ along the beam, namely the IBC signature \cite{Thomsen2019} (see the schematic at the bottom of the workflow in Fig.~\ref{fig:fig04}(b)). To this end, a source transfer function (STF) $h^{-1}_{LS/RS}(t)$ for both the left ($LS$) and right ($RS$) actuators must be found which accounts for both the mechanical response of the actuator ($m(t)$) and propagation ($p(t)$) along the beam to the measurement point at a distance $x_d$ to the source. The left and right ends of the beam were excited via the actuators and the STF determined using a measurement at a distance of $x_d = 0.1$ m from the respective ends of the beam. We then isolated the propagating direct wave ($v_{d}(x_{LS,RS},t)$), which must be clearly identifiable after wavefield separation, measuring at $x_{LS}=0.1$ m and $x_{RS}=0.9$ m respectively. The STF $h^{-1}_{LS/RS}(t)$ is eventually calculated by optimally matching the isolated direct wave to the original source signal ($s(t)$) applied to the actuator in a least-squares sense. A reliable estimation of the STF allows for an implementation of the IBC without any further assumptions about the medium required, as, e.g., propagation speeds or f-k relation. Thus, the method is fully data driven. 
Finally, the polarity reversed total reflected right- ($-r(x,t)_{refl}$) and leftward ($-l(x,t)_{refl}$) propagating wavefields were convolved with the STF $h^{-1}_{LS/RS}(t)$ and injected at the ends of the beam in the iterative and alternating manner described by Case 3. Resulting in the desired experimental state with the boundary reflections removed as shown in Fig.~\ref{fig:fig05}(b). This workflow is summarized by the flow chart depicted in Fig.~\ref{fig:fig04}(b). The final IBC signature shown by the black traces on the left and right end of Fig.~\ref{fig:fig05}(b) removed the first order reflection, as well as all higher order reflected events at the domain ends up to a time of 25 ms. The corresponding normalized spatial frequency distribution along the structure is depicted in Fig. \ref{fig:fig05}(d). We observe that the peak in the spatial frequency content at 585 Hz present in Fig. \ref{fig:fig05}(c) is absent due to the removal of the boundary reflections. Overall the induced band gap between 200 and 680 Hz is also more clearly discernible. Finally, an analysis of the frequency response measured at the tip of wing 13 also shows a mitigation of the strong peak at 805 Hz (see Fig.~\ref{fig:fig06}(b)), making it clear that the wing resonates at 220 Hz. 

\begin{figure}[ht]
\begin{center}
\includegraphics[]{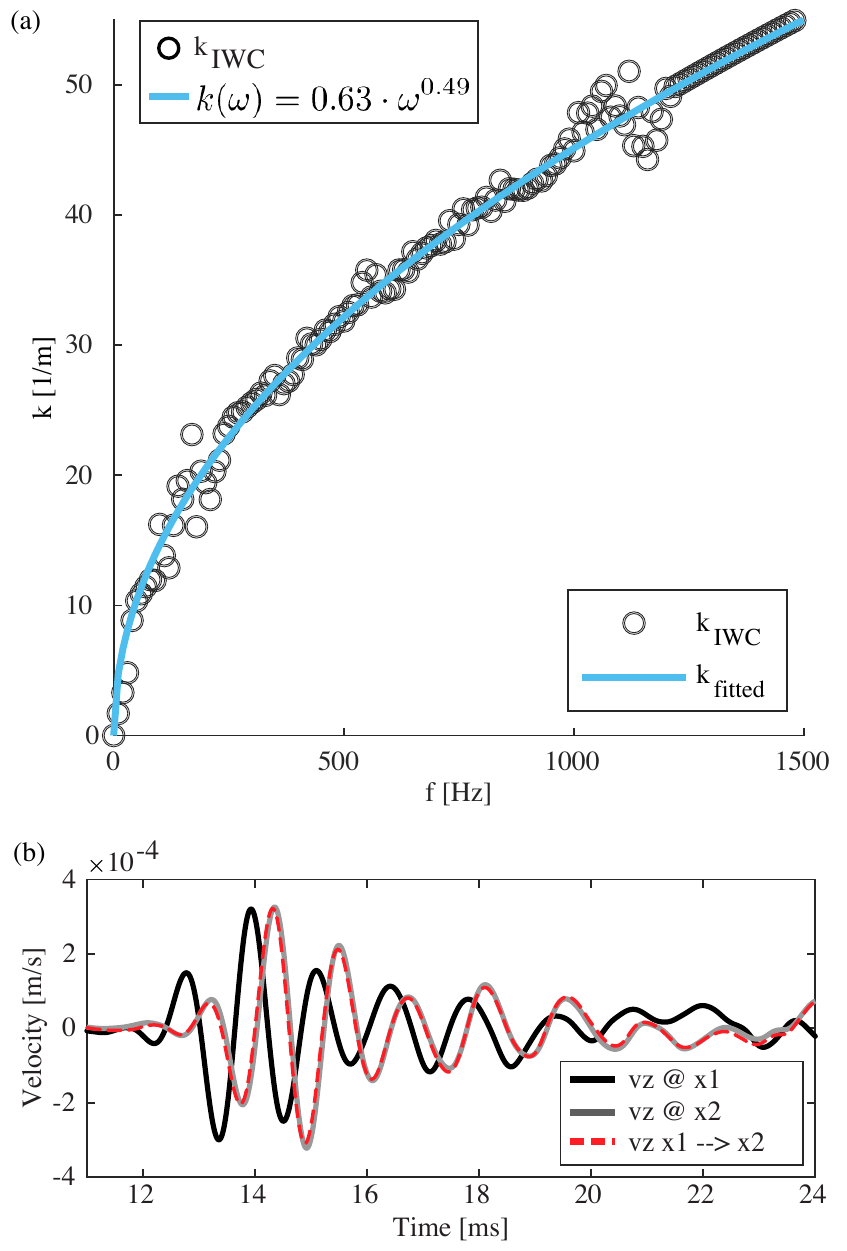}
\caption{(a) Dispersion relation as determined via the IWC method (black circles) and the fitted exponential function used for wavefield repositioning. (b) Forward redatuming of the measurement at $x_1 = 0.7$ m (black line) by 20 cm (dashed, red line) compared to the measurememt at $x_2 = 0.9$~m (gray line). \label{fig:fig07}}
\end{center}
\end{figure}

\subsection{Step 2: introducing an experimental virtual periodicity}

Once the reflections were removed, we experimentally introduced a virtual periodic boundary condition. To this end, the rightward propagating wavefield, impinging on the right end of the beam after passing through the graded area, is isolated and re-introducing on the left end of the beam. Since the wavefield could not be measured directly at the beam ends due to the presence of the actuators, it was measured at $x_1 = 0.85$ m and forward extrapolated to the right end of the beam at $x_2 = 1.0$ m before re-injection on the left side of the beam. Assuming plane wave propagation, the wavefield at $x_2$ can be expressed as $V(x_{2},\omega) = V(x_{1},\omega)e^{ikd}$ in the Fourier domain, where $k$ is the flexural wavenumber, $d = x_2 - x_1$ the propagation distance and $V(x_{1},\omega)$ the wavefield measured at $x_1$. Thus, the experimental dispersion relation had to be determined.

The required dispersion relation was obtained by adopting the Inhomogeneous Wave Correlation (IWC) method \cite{van2018measuring} which allows for calculation of the wave number by maximizing the correlation between a theoretical inhomogeneous running wave and the spatial response of the target over short measurement distances. We actuated the flexural $A_0$ Lamb mode on the right end of the beam and used the IWC method to determine the fk-relation of the host beam after the grading, i.e. for $x>0.6$ m. Figure \ref{fig:fig07}(a) shows the resulting dispersion relation, indicated by the black (dark) circles in the range of 0 - 1200 Hz. Assuming the $A_0$ mode can be described by an exponential power law function $k(\omega) = a\omega^b+c$, with $\omega=2 \pi f$, we used a least squares solver to determine a continues fk-relation in the frequency area of interest, indicated by the (light) blue line in Fig. \ref{fig:fig07}(a). To verify the result, we forward propagated (re-positioned) the wavefield measured at $x_1 = 0.7$ m (black line in Fig. \ref{fig:fig07}(b)) by $d = 0.2$ m to $x_2 = 0.9$ m (red dashed line in Fig. \ref{fig:fig07}(b)) and compared it to the measured wavefield at $x_2$ (gray line in Fig. \ref{fig:fig07}(b)) observing a good match.

\begin{figure}[ht!]
\begin{center}
\includegraphics[]{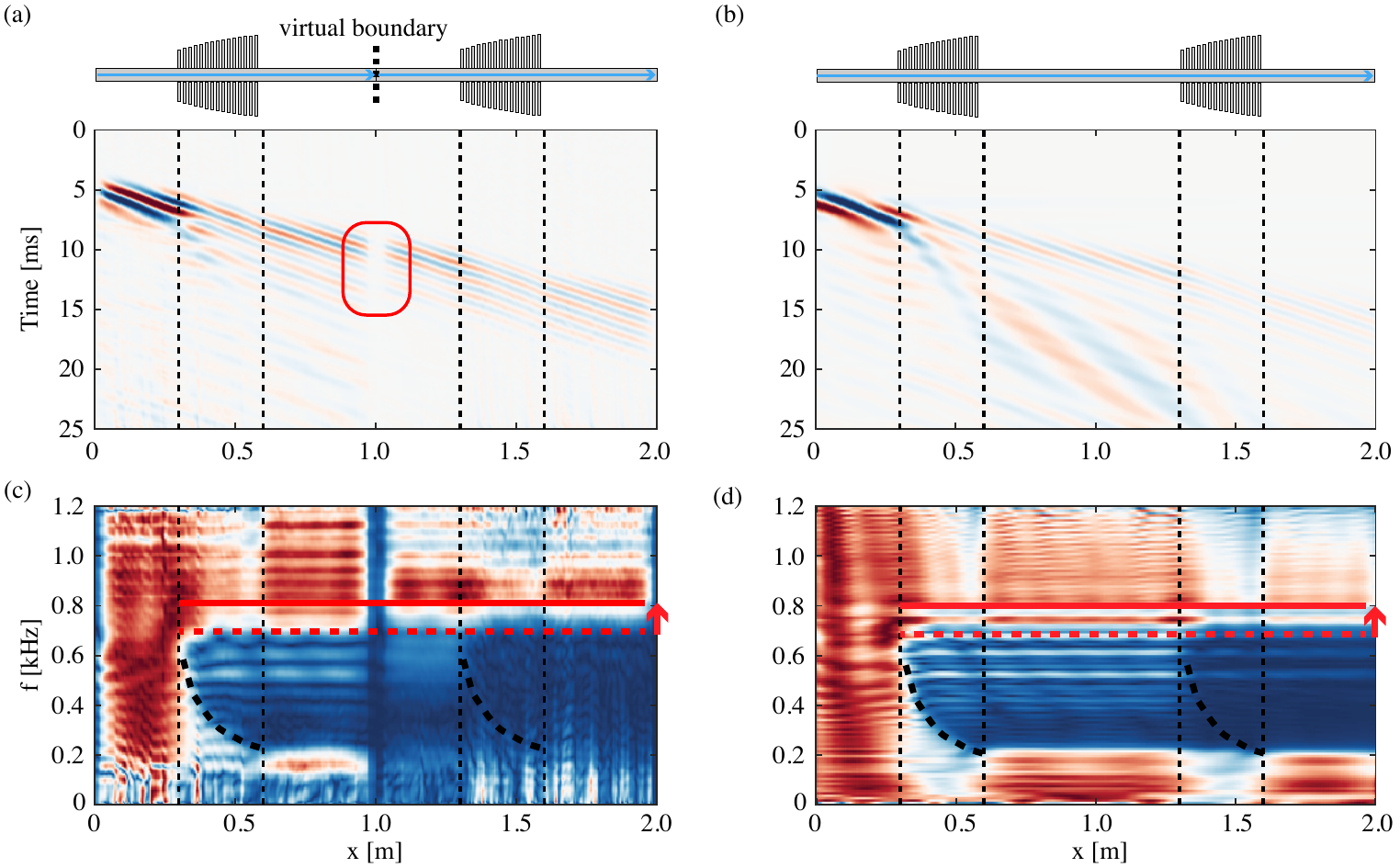}
\caption{Time-distance representation of the right going out-of-plane velocity measured along the beam. (a) Experimental data, the red box indicates where the rightward propagating wavefield is reintroduced on the left end, making it appear like the wave is propagating through a continuous medium with two sequential graded areas. (b) Numerical simulation of two sequential graded supercells and the resulting rightward wavefield propagation. (c \& d) The respective normalized spatial frequency representation of (a) and (b). The dashed red line indicates the upper end of the band gap after a single pass through the supercell. After the second pass through, a widening of the band gap and convergence to the analytical solution (solid red line) is observable in both the experimental and numerical results. \label{fig:fig08}}
\end{center}
\end{figure}

Finally, the right-going wavefield measured at $0.85$ m with the IBC active (see Fig.~\ref{fig:fig07}(b)) was forward propagated from $0.85$ to $1.0$ m. To utilize the STF of the left actuator, where the wavefield was re-introduced, the wavefield is extrapolated an additional $0.1$ m, such that its virtual recording position matches $x_{LS} = 0.1$ m, the measurement point for which $h^{-1}_{LS}(t)$ was determined. After applying the STF, the re-positioned, rightward propagating wave was re-introduced into the beam on the left end. The newly arising reflections were again cancelled by iteratively constructing a secondary IBC on both ends of the beam. 
The schematic of the prototype geometry in Fig.~\ref{fig:fig08}(a) indicates that the right end of the beam acts as a virtually periodic boundary as a result. The time-distance plot in the same Figure depicts the rightward propagating wavefield of the experimentally formed periodic boundary and Fig.~\ref{fig:fig08}(c) its normalized spatial frequency representation.

A numerical \textsc{COMSOL}$^{\circledR}$ \cite{COMSOL} simulation was used to verify the observed experimental phenomena (see Fig.~\ref{fig:fig08}(b\&d). The supercell of the numerical model, consisting of the 1 m long host beam and graded area, matches the experimental prototype in its geometry. The material properties of the prototype were approximated by a Young modulus of 69 GPa, density of 2710 $\mathrm{g/m^3}$ and Poisson ratio of 0.33. The supercell was replicated once, as indicated by the schematic in Fig.~\ref{fig:fig08}(b), to represent the periodic metamaterial. Comparing the experimental and numerical data, the band gap is deepened in both cases as the wave passes through the grading a second time. Furthermore, we found that, the induced periodicity recovers the analytical solution of the periodic boundary condition (further details can be found in the Appendix). After the first pass through the supercell, the upper end of the band gap (dashed red line) converges to the analytical solution (solid red line) in both the experimental and numerical results (Fig.~\ref{fig:fig08}(c\&d)). 

Important to note is that by re-introducing the wavefield exiting on the right back at the left end, the back scattered interactions between the virtually periodic graded metamaterial were not accounted for in this experimental realization. These interactions were however, included in the COMSOL simulation as the numerical model actually consists of two graded areas. This is where the differences in frequency content observable between $0.6$ - $1.3$ m in Fig.~\ref{fig:fig08}(c\&d) might originate from. One additional difference observable is the missing energy content below 200 Hz in the virtually periodic experimental realization. Taking a closer look at the rightward propagating wavefields in Fig.~\ref{fig:fig08}(a\&b), one may notice that the slower, low frequency portion of the propagating wave clearly visible in the numerical simulation, is absent in the experimental realisation during the second pass through the metamaterial (indicated by the red circle in Fig.~\ref{fig:fig08}(a)). 

\section{Discussion}

The immersive boundary method we have implemented paves avenues which require further exploration. First, the experimentally induced virtual periodicity in its current form accounts for the rightward propagating waves, i.e. one can observe waves passing through sequential supercells of the grading and the resulting effect. However, the leftward propagating waves, i.e., energy scattered at the graded resonator profile, are not accounted for. To include these interactions, construction of another virtual link is required. The second link would need to incorporate the leftward propagating wavefield reflected off the graded area after the first pass through. This wavefield would then be re-injected at the right end of the structure. Second, by applying the IBC's far away from the graded area, we are introducing the periodicity based on the propagating wavefield but not the evanescent, non-propagating component. However, the evanescent, short-range interactions between the graded supercells could play an important role in the band gap creation \cite{Lemoult2010,Lemoult2011}. In the workflow presented, the direct wave must be properly separated from the reflections to determine the STF. Hence, at least one wavelength before the scattering is required. Moving the actuators and, thus, IBC closer to the graded area is theoretically possible, but would require a different approach to separate the wavefields. Finally, after the introduction of the virtual periodicity in Fig.~\ref{fig:fig08}(a\&c), missing frequency content $<$200 Hz could be observed. When first applying the periodic signal a new set of IBCs had to be constructed. To preserve the periodic signal in the process, it had to be separated from the new, rightward propagating, unwanted reflections occurring at the left end through windowing. However, the much slower, low frequency component of the periodic signal was indistinguishable from these reflections. Consequently, the $<$200 Hz portion of the periodic signal was erroneously included in the first iteration of the new IBC and removed through its application. Instead of utilizing visual windowing, the two wavefield components could be separated by subtracting the predicted periodic signal from the full rightward propagating wavefield. 

The examples presented in this work showcase propagating waves with limited spatial and temporal width, i.e. travelling wave packages. In this first realization such propagation was desirable, as it made identification of the direct wave and its first to higher order reflections feasible. However, our proposed workflow could also be applied to periodic signals, such as continuous sinusoidal excitation of the metamaterial. In fact, the signature of the propagating wavefield confined to the first 30 cm of the metamaterial, i.e. before the grading, observable in Fig.~\ref{fig:fig07}(a), has strong similarities to standing waves emerging during periodic excitation. Indeed, the signature of the designed IBC on the left end of the beam, as depicted in Fig.~\ref{fig:fig07}(b), resembles a continuous sinusoidal excitation. Hence, by manipulating the boundary conditions, our methodology could be used to, e.g., transform standing waves to travelling waves and enrich the data sampled along the metamaterial under investigation. The possible advantages of such an experimental tool were recently discussed in an acoustic impedance tube \cite{Vered2021}. 

Thanks to the possibilities offered by boundless metamaterial experimentation, a real time implementation of the method presented here is highly desirable. However, a real time realization of elastic IWE is challenging and its feasibility strongly depends on the experimental conditions. Whereas acoustic implementations of IWE have been achieved in real time \cite{Becker2021}, the much larger wave speeds in solids, when compared to air or fluids, are the strongest limiting factor. Enough time is needed to predict the wavefield incident on the boundary where the desired manipulation via IWE is to take place. If the location where the wavefield is recorded and the IBC applied are the same, a real time implementation is not possible. Nevertheless, in a set-up similar to the one presented here, where recordings along the wave propagation path can be made, a real time implementation could be possible with sufficiently fast hardware. In any case, it is necessary to extrapolate the wavefield from the recording to the emitting boundary and, thus, a prior characterization of a homogeneous portion of the medium under investigation is required. The feasibility could be further increased by utilizing manufactured materials with much slower wave speeds, such as foams or plastics. 

We would like to point out that our proposed methodology has similarities to "surface related multiple removal" \cite{Abma2005}, commonly applied in the field of geophysics. However, our method does not rely on the prediction of the multiple reflections occurring at the domain boundaries but rather is fully data driven, assuming only linear wave propagation.

Finally, full three dimensional, 3-component elastic IWE, as proposed by \citet{Vasmel2016b} and \citet{Thomsen2019}, has yet to be realized experimentally. To this end, \citet{Thomsen2021} proposed a numerical wavefield injection based methodology to retrieve the primary incident wavefield when only recordings of the three-component particle velocity at the free surface are available. However, the method requires a full characterization of the physical experimental domain. Recently, \citet{Li2022} bypassed this, in most experimental cases unfeasible, requirement by incorporating an internal absorbing boundary in the wavefield injection scheme, resulting in the retrieval of the \textit{full} incident and reflected wavefields. Consequently, this wavefield injection scheme could be combined with our proposed iterative workflow to utilize the \textit{full} reflected wavefield in designing the IBC instead of only the primary incident wavefield. 

\section{Conclusion}

We introduced a novel approach to the experimental investigation and validation of metamaterial prototypes. Providing a fully data driven workflow to implement active boundary control in the presence of strongly dispersive wave propagation and internal scattering. Through the design of an immersive boundary condition (IBC) reflection suppression, virtual periodicity, and the introduction of fictitious boundaries is enabled. The IBC is constructed iteratively by injection of the polarity reversed full wavefield reflected at the target boundaries. Higher order reflections are removed step-by-step by adapting the IBC during each experimental iteration. Our ad-hoc workflow circumvents the requirement of a perfect free surface boundary to act on with the IBC, as defined by elastic IWE theory \cite{Thomsen2019,Becker2020}, purely working with the reflected wavefield and requires no knowledge of the medium under investigation. Thus, it permits any experimental conditions, such as clamped boundaries or those deviating from prefect free surfaces due to the influence of attached actuators. In a first experimental proof of concept a virtual periodic boundary condition was created in an elastic metamaterial by re-introducing waves leaving the structure on one end back into the opposite end. As a result, the experimental band gap observed converged to the analytical solution. This experimental realization of a ``Bloch-like'' geometric periodicity has the potential to facilitate the study of metamaterials without the need to physically realize large structures in the laboratory. Instead only one to a few unit cells are required. Boundless experimentation can thus act as a powerful tool in future metamaterial research.

\begin{acknowledgments}
This research was supported by the Horizon 2020 FET-proactive project METAVEH under the grant agreement 952039 and the ERC grant 694407 - MATRIX.
\end{acknowledgments}

\appendix

\section{Appendixes}

As observable in Fig.~\ref{fig:fig08}(c), the experimental band gap widens when the virtual periodicity is introduced. This phenomenon can be understood as the experimental realization converging to the analytical solution. 

Recently \citet{Pu2022} proposed a multi scatter formulation to analytically describe the out-of-plane component $w(x)$ of a wave at position $x$ propagating along a host beam and interacting with $N$ resonators as:

\begin{equation}
    w(x)=w_0(x)+\sum_{n=1}^N Q_n G_w(x-x_n).
        \label{eq:A1}
\end{equation}

Here $w_0(x)$ represents the wave field at position $x$ incident at the base of $N$ parasitic wing pairs. The term $Q_n$ describes the amplitude of the uniform normal stress at the base of the $n$-th pair of wings attached to the host beam. $G_w$ is the related out-of-plane displacement Green's function \cite{graff2012wave}. Together these terms compute the scattered wave field of the $N$ parasitic wing pairs. The normal stress amplitude $Q_n$ can be found from the mechanical impedance via:

\begin{equation}
    Q_n=\frac{2M_n\omega_{rn}^2\omega^2}{S_n(\omega_{rn}^2-\omega^2)}w(x_n)=Z_nw(x_n),
        \label{eq:A2}
\end{equation}
where $\omega_{rn}=\sqrt{K_n/M_n}$ is the resonant frequency of the $n$-th pair of parasitic wings. Together with the mechanical impedance $Z_n$, the amplitude $Q_n$ can be computed knowing the vertical displacement $w(x_n)$ at the base of the $n$-th parasitic wing pair. Thus, given a incident wavefield $w_0(x)$, the stress amplitudes can be computed and the total wavefield obtained using Eq.~\ref{eq:A1}. 

Assuming a null incident field with $w_0=0$, the eigenstates of the system can be obtained via Eq.~\ref{eq:A1} and Eq.~\ref{eq:A2} in matrix form from the eigenvalue problem:

\begin{equation}
    \mathbf{AX}=\mathbf{0},
    \label{eq:A3}
\end{equation}

with:

\begin{equation}
\mathbf{A}=\left[\begin{array}{cccc}
Z_1^{-1}-G_w(0) & -G_w\left(x_1-x_2\right) & \cdots & -G_w\left(x_1-x_N\right) \\
-G_w\left(x_2-x_1\right) & Z_2^{-1}-G_w(0) & \cdots & -G_w\left(x_2-x_N\right) \\
\vdots & \vdots & \ddots & \vdots \\
-G_w\left(x_N-x_1\right) & -G_w\left(x_N-x_2\right) & \cdots & Z_N^{-1}-G_w(0)
\end{array}\right], \mathbf{X}=\left[\begin{array}{c}
Q_1 \\
Q_2 \\
\vdots \\
Q_N
\end{array}\right].
    \label{eq:A4}
\end{equation}

The virtual periodicity introduced via our proposed method only accounts for rightward propagating waves and neglects the interactions between supercells. A supercell describes one element of the host beam plus graded parasitic wings as illustrated in Fig.~\ref{fig:fig08}(a). Consequently, this extended eigenvalue problem can be formulated as:

\begin{equation}
    \mathbf{A_EX_E}=\mathbf{0},
        \label{eq:A5}
\end{equation}

with:

\begin{equation}
\mathbf{A_E}=\left[\begin{array}{cccc}
\mathbf{A_1} & & & \\
& \mathbf{A_2} & & \\
& & \ddots &\\
& & & \mathbf{A_N}
\end{array}\right], \mathbf{X_E}=\left[\begin{array}{c}
\mathbf{X_1} \\
\mathbf{X_2} \\
\vdots \\
\mathbf{X_N}
\end{array}\right],
    \label{eq:A6}
\end{equation}

where $\mathbf{A_1}$ to $\mathbf{A_N}$ describe the periodically arranged supercells. Comparing Eq.~\ref{eq:A6} to Eq.~\ref{eq:A4} this configuration has the same eigenvalues corresponding to the bandgap ranges of a singular supercell of the metamaterial. Thus, by introducing the virtual, geometric periodicity experimentally, one converges to the analytical solution of the eigenvalue problem determining the metamaterials bandgap. 

\bibliography{References}% Produces the bibliography via BibTeX.

%apsrev4-2.bst 2019-01-14 (MD) hand-edited version of apsrev4-1.bst
%Control: key (0)
%Control: author (8) initials jnrlst
%Control: editor formatted (1) identically to author
%Control: production of article title (0) allowed
%Control: page (0) single
%Control: year (1) truncated
%Control: production of eprint (0) enabled
\begin{thebibliography}{40}%
\makeatletter
\providecommand \@ifxundefined [1]{%
 \@ifx{#1\undefined}
}%
\providecommand \@ifnum [1]{%
 \ifnum #1\expandafter \@firstoftwo
 \else \expandafter \@secondoftwo
 \fi
}%
\providecommand \@ifx [1]{%
 \ifx #1\expandafter \@firstoftwo
 \else \expandafter \@secondoftwo
 \fi
}%
\providecommand \natexlab [1]{#1}%
\providecommand \enquote  [1]{``#1''}%
\providecommand \bibnamefont  [1]{#1}%
\providecommand \bibfnamefont [1]{#1}%
\providecommand \citenamefont [1]{#1}%
\providecommand \href@noop [0]{\@secondoftwo}%
\providecommand \href [0]{\begingroup \@sanitize@url \@href}%
\providecommand \@href[1]{\@@startlink{#1}\@@href}%
\providecommand \@@href[1]{\endgroup#1\@@endlink}%
\providecommand \@sanitize@url [0]{\catcode `\\12\catcode `\$12\catcode
  `\&12\catcode `\#12\catcode `\^12\catcode `\_12\catcode `\%12\relax}%
\providecommand \@@startlink[1]{}%
\providecommand \@@endlink[0]{}%
\providecommand \url  [0]{\begingroup\@sanitize@url \@url }%
\providecommand \@url [1]{\endgroup\@href {#1}{\urlprefix }}%
\providecommand \urlprefix  [0]{URL }%
\providecommand \Eprint [0]{\href }%
\providecommand \doibase [0]{https://doi.org/}%
\providecommand \selectlanguage [0]{\@gobble}%
\providecommand \bibinfo  [0]{\@secondoftwo}%
\providecommand \bibfield  [0]{\@secondoftwo}%
\providecommand \translation [1]{[#1]}%
\providecommand \BibitemOpen [0]{}%
\providecommand \bibitemStop [0]{}%
\providecommand \bibitemNoStop [0]{.\EOS\space}%
\providecommand \EOS [0]{\spacefactor3000\relax}%
\providecommand \BibitemShut  [1]{\csname bibitem#1\endcsname}%
\let\auto@bib@innerbib\@empty
%</preamble>
\bibitem [{\citenamefont {Bertoldi}\ \emph {et~al.}(2010)\citenamefont
  {Bertoldi}, \citenamefont {Reis}, \citenamefont {Willshaw},\ and\
  \citenamefont {Mullin}}]{Bertoldi2010}%
  \BibitemOpen
  \bibfield  {author} {\bibinfo {author} {\bibfnamefont {K.}~\bibnamefont
  {Bertoldi}}, \bibinfo {author} {\bibfnamefont {P.~M.}\ \bibnamefont {Reis}},
  \bibinfo {author} {\bibfnamefont {S.}~\bibnamefont {Willshaw}},\ and\
  \bibinfo {author} {\bibfnamefont {T.}~\bibnamefont {Mullin}},\ }\bibfield
  {title} {\bibinfo {title} {Negative poisson's ratio behavior induced by an
  elastic instability},\ }\href@noop {} {\bibfield  {journal} {\bibinfo
  {journal} {Advanced materials}\ }\textbf {\bibinfo {volume} {22}},\ \bibinfo
  {pages} {361} (\bibinfo {year} {2010})}\BibitemShut {NoStop}%
\bibitem [{\citenamefont {Zheng}\ \emph {et~al.}(2014)\citenamefont {Zheng},
  \citenamefont {Lee}, \citenamefont {Weisgraber}, \citenamefont {Shusteff},
  \citenamefont {DeOtte}, \citenamefont {Duoss}, \citenamefont {Kuntz},
  \citenamefont {Biener}, \citenamefont {Ge}, \citenamefont {Jackson} \emph
  {et~al.}}]{Zheng2014}%
  \BibitemOpen
  \bibfield  {author} {\bibinfo {author} {\bibfnamefont {X.}~\bibnamefont
  {Zheng}}, \bibinfo {author} {\bibfnamefont {H.}~\bibnamefont {Lee}}, \bibinfo
  {author} {\bibfnamefont {T.~H.}\ \bibnamefont {Weisgraber}}, \bibinfo
  {author} {\bibfnamefont {M.}~\bibnamefont {Shusteff}}, \bibinfo {author}
  {\bibfnamefont {J.}~\bibnamefont {DeOtte}}, \bibinfo {author} {\bibfnamefont
  {E.~B.}\ \bibnamefont {Duoss}}, \bibinfo {author} {\bibfnamefont {J.~D.}\
  \bibnamefont {Kuntz}}, \bibinfo {author} {\bibfnamefont {M.~M.}\ \bibnamefont
  {Biener}}, \bibinfo {author} {\bibfnamefont {Q.}~\bibnamefont {Ge}}, \bibinfo
  {author} {\bibfnamefont {J.~A.}\ \bibnamefont {Jackson}}, \emph {et~al.},\
  }\bibfield  {title} {\bibinfo {title} {Ultralight, ultrastiff mechanical
  metamaterials},\ }\href@noop {} {\bibfield  {journal} {\bibinfo  {journal}
  {Science}\ }\textbf {\bibinfo {volume} {344}},\ \bibinfo {pages} {1373}
  (\bibinfo {year} {2014})}\BibitemShut {NoStop}%
\bibitem [{\citenamefont {Wang}\ \emph {et~al.}(2021)\citenamefont {Wang},
  \citenamefont {Li}, \citenamefont {Hofmann}, \citenamefont {Andrade},\ and\
  \citenamefont {Daraio}}]{Wang2021}%
  \BibitemOpen
  \bibfield  {author} {\bibinfo {author} {\bibfnamefont {Y.}~\bibnamefont
  {Wang}}, \bibinfo {author} {\bibfnamefont {L.}~\bibnamefont {Li}}, \bibinfo
  {author} {\bibfnamefont {D.}~\bibnamefont {Hofmann}}, \bibinfo {author}
  {\bibfnamefont {J.~E.}\ \bibnamefont {Andrade}},\ and\ \bibinfo {author}
  {\bibfnamefont {C.}~\bibnamefont {Daraio}},\ }\bibfield  {title} {\bibinfo
  {title} {Structured fabrics with tunable mechanical properties},\ }\href@noop
  {} {\bibfield  {journal} {\bibinfo  {journal} {Nature}\ }\textbf {\bibinfo
  {volume} {596}},\ \bibinfo {pages} {238} (\bibinfo {year}
  {2021})}\BibitemShut {NoStop}%
\bibitem [{\citenamefont {Colombi}\ \emph {et~al.}(2016)\citenamefont
  {Colombi}, \citenamefont {Colquitt}, \citenamefont {Roux}, \citenamefont
  {Guenneau},\ and\ \citenamefont {Craster}}]{Colombi2016}%
  \BibitemOpen
  \bibfield  {author} {\bibinfo {author} {\bibfnamefont {A.}~\bibnamefont
  {Colombi}}, \bibinfo {author} {\bibfnamefont {D.}~\bibnamefont {Colquitt}},
  \bibinfo {author} {\bibfnamefont {P.}~\bibnamefont {Roux}}, \bibinfo {author}
  {\bibfnamefont {S.}~\bibnamefont {Guenneau}},\ and\ \bibinfo {author}
  {\bibfnamefont {R.~V.}\ \bibnamefont {Craster}},\ }\bibfield  {title}
  {\bibinfo {title} {A seismic metamaterial: The resonant metawedge},\
  }\href@noop {} {\bibfield  {journal} {\bibinfo  {journal} {Scientific
  reports}\ }\textbf {\bibinfo {volume} {6}},\ \bibinfo {pages} {1} (\bibinfo
  {year} {2016})}\BibitemShut {NoStop}%
\bibitem [{\citenamefont {Matlack}\ \emph {et~al.}(2016)\citenamefont
  {Matlack}, \citenamefont {Bauhofer}, \citenamefont {Kr{\"o}del},
  \citenamefont {Palermo},\ and\ \citenamefont {Daraio}}]{Matlack2016}%
  \BibitemOpen
  \bibfield  {author} {\bibinfo {author} {\bibfnamefont {K.~H.}\ \bibnamefont
  {Matlack}}, \bibinfo {author} {\bibfnamefont {A.}~\bibnamefont {Bauhofer}},
  \bibinfo {author} {\bibfnamefont {S.}~\bibnamefont {Kr{\"o}del}}, \bibinfo
  {author} {\bibfnamefont {A.}~\bibnamefont {Palermo}},\ and\ \bibinfo {author}
  {\bibfnamefont {C.}~\bibnamefont {Daraio}},\ }\bibfield  {title} {\bibinfo
  {title} {Composite 3{D}-printed metastructures for low-frequency and
  broadband vibration absorption},\ }\href@noop {} {\bibfield  {journal}
  {\bibinfo  {journal} {Proceedings of the National Academy of Sciences}\
  }\textbf {\bibinfo {volume} {113}},\ \bibinfo {pages} {8386} (\bibinfo {year}
  {2016})}\BibitemShut {NoStop}%
\bibitem [{\citenamefont {Sigalas}\ and\ \citenamefont
  {Economou}(1992)}]{sigalas1992elastic}%
  \BibitemOpen
  \bibfield  {author} {\bibinfo {author} {\bibfnamefont {M.~M.}\ \bibnamefont
  {Sigalas}}\ and\ \bibinfo {author} {\bibfnamefont {E.~N.}\ \bibnamefont
  {Economou}},\ }\bibfield  {title} {\bibinfo {title} {Elastic and acoustic
  wave band structure},\ }\href@noop {} {\bibfield  {journal} {\bibinfo
  {journal} {Journal of sound and vibration}\ }\textbf {\bibinfo {volume}
  {158}},\ \bibinfo {pages} {377} (\bibinfo {year} {1992})}\BibitemShut
  {NoStop}%
\bibitem [{\citenamefont {Liu}\ \emph {et~al.}(2000)\citenamefont {Liu},
  \citenamefont {Zhang}, \citenamefont {Mao}, \citenamefont {Zhu},
  \citenamefont {Yang}, \citenamefont {Chan},\ and\ \citenamefont
  {Sheng}}]{liu2000locally}%
  \BibitemOpen
  \bibfield  {author} {\bibinfo {author} {\bibfnamefont {Z.}~\bibnamefont
  {Liu}}, \bibinfo {author} {\bibfnamefont {X.}~\bibnamefont {Zhang}}, \bibinfo
  {author} {\bibfnamefont {Y.}~\bibnamefont {Mao}}, \bibinfo {author}
  {\bibfnamefont {Y.}~\bibnamefont {Zhu}}, \bibinfo {author} {\bibfnamefont
  {Z.}~\bibnamefont {Yang}}, \bibinfo {author} {\bibfnamefont {C.~T.}\
  \bibnamefont {Chan}},\ and\ \bibinfo {author} {\bibfnamefont
  {P.}~\bibnamefont {Sheng}},\ }\bibfield  {title} {\bibinfo {title} {Locally
  resonant sonic materials},\ }\href@noop {} {\bibfield  {journal} {\bibinfo
  {journal} {Science}\ }\textbf {\bibinfo {volume} {289}},\ \bibinfo {pages}
  {1734} (\bibinfo {year} {2000})}\BibitemShut {NoStop}%
\bibitem [{\citenamefont {Craster}\ and\ \citenamefont
  {Guenneau}(2012)}]{craster2012acoustic}%
  \BibitemOpen
  \bibfield  {author} {\bibinfo {author} {\bibfnamefont {R.~V.}\ \bibnamefont
  {Craster}}\ and\ \bibinfo {author} {\bibfnamefont {S.}~\bibnamefont
  {Guenneau}},\ }\href@noop {} {\emph {\bibinfo {title} {Acoustic
  metamaterials: Negative refraction, imaging, lensing and cloaking}}},\ Vol.\
  \bibinfo {volume} {166}\ (\bibinfo  {publisher} {Springer Science \& Business
  Media},\ \bibinfo {year} {2012})\BibitemShut {NoStop}%
\bibitem [{\citenamefont {De~Ponti}\ \emph
  {et~al.}(2020{\natexlab{a}})\citenamefont {De~Ponti}, \citenamefont
  {Colombi}, \citenamefont {Riva}, \citenamefont {Ardito}, \citenamefont
  {Braghin}, \citenamefont {Corigliano},\ and\ \citenamefont
  {Craster}}]{DePonti2020experimental}%
  \BibitemOpen
  \bibfield  {author} {\bibinfo {author} {\bibfnamefont {J.~M.}\ \bibnamefont
  {De~Ponti}}, \bibinfo {author} {\bibfnamefont {A.}~\bibnamefont {Colombi}},
  \bibinfo {author} {\bibfnamefont {E.}~\bibnamefont {Riva}}, \bibinfo {author}
  {\bibfnamefont {R.}~\bibnamefont {Ardito}}, \bibinfo {author} {\bibfnamefont
  {F.}~\bibnamefont {Braghin}}, \bibinfo {author} {\bibfnamefont
  {A.}~\bibnamefont {Corigliano}},\ and\ \bibinfo {author} {\bibfnamefont
  {R.~V.}\ \bibnamefont {Craster}},\ }\bibfield  {title} {\bibinfo {title}
  {Experimental investigation of amplification, via a mechanical delay-line, in
  a rainbow-based metamaterial for energy harvesting},\ }\href@noop {}
  {\bibfield  {journal} {\bibinfo  {journal} {Applied Physics Letters}\
  }\textbf {\bibinfo {volume} {117}},\ \bibinfo {pages} {143902} (\bibinfo
  {year} {2020}{\natexlab{a}})}\BibitemShut {NoStop}%
\bibitem [{\citenamefont {De~Ponti}\ \emph
  {et~al.}(2020{\natexlab{b}})\citenamefont {De~Ponti}, \citenamefont
  {Colombi}, \citenamefont {Ardito}, \citenamefont {Braghin}, \citenamefont
  {Corigliano},\ and\ \citenamefont {Craster}}]{DePonti2020graded}%
  \BibitemOpen
  \bibfield  {author} {\bibinfo {author} {\bibfnamefont {J.~M.}\ \bibnamefont
  {De~Ponti}}, \bibinfo {author} {\bibfnamefont {A.}~\bibnamefont {Colombi}},
  \bibinfo {author} {\bibfnamefont {R.}~\bibnamefont {Ardito}}, \bibinfo
  {author} {\bibfnamefont {F.}~\bibnamefont {Braghin}}, \bibinfo {author}
  {\bibfnamefont {A.}~\bibnamefont {Corigliano}},\ and\ \bibinfo {author}
  {\bibfnamefont {R.~V.}\ \bibnamefont {Craster}},\ }\bibfield  {title}
  {\bibinfo {title} {Graded elastic metasurface for enhanced energy
  harvesting},\ }\href@noop {} {\bibfield  {journal} {\bibinfo  {journal} {New
  Journal of Physics}\ }\textbf {\bibinfo {volume} {22}},\ \bibinfo {pages}
  {013013} (\bibinfo {year} {2020}{\natexlab{b}})}\BibitemShut {NoStop}%
\bibitem [{\citenamefont {De~Ponti}\ \emph
  {et~al.}(2021{\natexlab{a}})\citenamefont {De~Ponti}, \citenamefont {Iorio},
  \citenamefont {Riva}, \citenamefont {Braghin}, \citenamefont {Corigliano},\
  and\ \citenamefont {Ardito}}]{DePonti2021enhanced}%
  \BibitemOpen
  \bibfield  {author} {\bibinfo {author} {\bibfnamefont {J.~M.}\ \bibnamefont
  {De~Ponti}}, \bibinfo {author} {\bibfnamefont {L.}~\bibnamefont {Iorio}},
  \bibinfo {author} {\bibfnamefont {E.}~\bibnamefont {Riva}}, \bibinfo {author}
  {\bibfnamefont {F.}~\bibnamefont {Braghin}}, \bibinfo {author} {\bibfnamefont
  {A.}~\bibnamefont {Corigliano}},\ and\ \bibinfo {author} {\bibfnamefont
  {R.}~\bibnamefont {Ardito}},\ }\bibfield  {title} {\bibinfo {title} {Enhanced
  energy harvesting of flexural waves in elastic beams by bending mode of
  graded resonators},\ }\href@noop {} {\bibfield  {journal} {\bibinfo
  {journal} {Frontiers in Materials}\ }\textbf {\bibinfo {volume} {8}}
  (\bibinfo {year} {2021}{\natexlab{a}})}\BibitemShut {NoStop}%
\bibitem [{\citenamefont {Zhao}\ \emph {et~al.}(2022)\citenamefont {Zhao},
  \citenamefont {Thomsen}, \citenamefont {{De Ponti}}, \citenamefont {Riva},
  \citenamefont {{Van Damme}}, \citenamefont {Bergamini}, \citenamefont
  {Chatzi},\ and\ \citenamefont {Colombi}}]{Bao2022}%
  \BibitemOpen
  \bibfield  {author} {\bibinfo {author} {\bibfnamefont {B.}~\bibnamefont
  {Zhao}}, \bibinfo {author} {\bibfnamefont {H.~R.}\ \bibnamefont {Thomsen}},
  \bibinfo {author} {\bibfnamefont {J.~M.}\ \bibnamefont {{De Ponti}}},
  \bibinfo {author} {\bibfnamefont {E.}~\bibnamefont {Riva}}, \bibinfo {author}
  {\bibfnamefont {B.}~\bibnamefont {{Van Damme}}}, \bibinfo {author}
  {\bibfnamefont {A.}~\bibnamefont {Bergamini}}, \bibinfo {author}
  {\bibfnamefont {E.}~\bibnamefont {Chatzi}},\ and\ \bibinfo {author}
  {\bibfnamefont {A.}~\bibnamefont {Colombi}},\ }\bibfield  {title} {\bibinfo
  {title} {A graded metamaterial for broadband and high-capability
  piezoelectric energy harvesting},\ }\href
  {https://doi.org/https://doi.org/10.1016/j.enconman.2022.116056} {\bibfield
  {journal} {\bibinfo  {journal} {Energy Conversion and Management}\ }\textbf
  {\bibinfo {volume} {269}},\ \bibinfo {pages} {116056} (\bibinfo {year}
  {2022})}\BibitemShut {NoStop}%
\bibitem [{\citenamefont {Komatitsch}\ and\ \citenamefont
  {Martin}(2007)}]{Komatitsch2007}%
  \BibitemOpen
  \bibfield  {author} {\bibinfo {author} {\bibfnamefont {D.}~\bibnamefont
  {Komatitsch}}\ and\ \bibinfo {author} {\bibfnamefont {R.}~\bibnamefont
  {Martin}},\ }\bibfield  {title} {\bibinfo {title} {An unsplit convolutional
  perfectly matched layer improved at grazing incidence for the seismic wave
  equation},\ }\href {https://doi.org/10.1190/1.2757586} {\bibfield  {journal}
  {\bibinfo  {journal} {Geophysics}\ }\textbf {\bibinfo {volume} {72}},\
  \bibinfo {pages} {SM155} (\bibinfo {year} {2007})}\BibitemShut {NoStop}%
\bibitem [{\citenamefont {Rajagopal}\ \emph {et~al.}(2012)\citenamefont
  {Rajagopal}, \citenamefont {Drozdz}, \citenamefont {Skelton}, \citenamefont
  {Lowe},\ and\ \citenamefont {Craster}}]{RAJAGOPAL201230}%
  \BibitemOpen
  \bibfield  {author} {\bibinfo {author} {\bibfnamefont {P.}~\bibnamefont
  {Rajagopal}}, \bibinfo {author} {\bibfnamefont {M.}~\bibnamefont {Drozdz}},
  \bibinfo {author} {\bibfnamefont {E.~A.}\ \bibnamefont {Skelton}}, \bibinfo
  {author} {\bibfnamefont {M.~J.}\ \bibnamefont {Lowe}},\ and\ \bibinfo
  {author} {\bibfnamefont {R.~V.}\ \bibnamefont {Craster}},\ }\bibfield
  {title} {\bibinfo {title} {On the use of absorbing layers to simulate the
  propagation of elastic waves in unbounded isotropic media using commercially
  available finite element packages},\ }\href
  {https://doi.org/https://doi.org/10.1016/j.ndteint.2012.04.001} {\bibfield
  {journal} {\bibinfo  {journal} {NDT \& E International}\ }\textbf {\bibinfo
  {volume} {51}},\ \bibinfo {pages} {30} (\bibinfo {year} {2012})}\BibitemShut
  {NoStop}%
\bibitem [{\citenamefont {Mironov}(1988)}]{Mironov1988}%
  \BibitemOpen
  \bibfield  {author} {\bibinfo {author} {\bibfnamefont {M.}~\bibnamefont
  {Mironov}},\ }\href@noop {} {\bibinfo {title} {Propagation of a flexural wave
  in a plate whose thickness decreases smoothly to zero in a finite interval}}
  (\bibinfo {year} {1988})\BibitemShut {NoStop}%
\bibitem [{\citenamefont {O’Boy}\ \emph {et~al.}(2010)\citenamefont
  {O’Boy}, \citenamefont {Krylov},\ and\ \citenamefont
  {Kralovic}}]{Oboy2010}%
  \BibitemOpen
  \bibfield  {author} {\bibinfo {author} {\bibfnamefont {D.}~\bibnamefont
  {O’Boy}}, \bibinfo {author} {\bibfnamefont {V.}~\bibnamefont {Krylov}},\
  and\ \bibinfo {author} {\bibfnamefont {V.}~\bibnamefont {Kralovic}},\
  }\bibfield  {title} {\bibinfo {title} {Damping of flexural vibrations in
  rectangular plates using the acoustic black hole effect},\ }\href
  {https://doi.org/https://doi.org/10.1016/j.jsv.2010.05.019} {\bibfield
  {journal} {\bibinfo  {journal} {Journal of Sound and Vibration}\ }\textbf
  {\bibinfo {volume} {329}},\ \bibinfo {pages} {4672} (\bibinfo {year}
  {2010})}\BibitemShut {NoStop}%
\bibitem [{\citenamefont {Georgiev}\ \emph {et~al.}(2011)\citenamefont
  {Georgiev}, \citenamefont {Cuenca}, \citenamefont {Gautier}, \citenamefont
  {Simon},\ and\ \citenamefont {Krylov}}]{Georgiev2011}%
  \BibitemOpen
  \bibfield  {author} {\bibinfo {author} {\bibfnamefont {V.}~\bibnamefont
  {Georgiev}}, \bibinfo {author} {\bibfnamefont {J.}~\bibnamefont {Cuenca}},
  \bibinfo {author} {\bibfnamefont {F.}~\bibnamefont {Gautier}}, \bibinfo
  {author} {\bibfnamefont {L.}~\bibnamefont {Simon}},\ and\ \bibinfo {author}
  {\bibfnamefont {V.}~\bibnamefont {Krylov}},\ }\bibfield  {title} {\bibinfo
  {title} {Damping of structural vibrations in beams and elliptical plates
  using the acoustic black hole effect},\ }\href
  {https://doi.org/https://doi.org/10.1016/j.jsv.2010.12.001} {\bibfield
  {journal} {\bibinfo  {journal} {Journal of Sound and Vibration}\ }\textbf
  {\bibinfo {volume} {330}},\ \bibinfo {pages} {2497} (\bibinfo {year}
  {2011})}\BibitemShut {NoStop}%
\bibitem [{\citenamefont {Vemula}\ \emph {et~al.}(1996)\citenamefont {Vemula},
  \citenamefont {Norris},\ and\ \citenamefont {Cody}}]{Vemula1996}%
  \BibitemOpen
  \bibfield  {author} {\bibinfo {author} {\bibfnamefont {C.}~\bibnamefont
  {Vemula}}, \bibinfo {author} {\bibfnamefont {A.}~\bibnamefont {Norris}},\
  and\ \bibinfo {author} {\bibfnamefont {G.}~\bibnamefont {Cody}},\ }\bibfield
  {title} {\bibinfo {title} {Attenuation of waves in plates and bares using a
  graded impedance interface at edges},\ }\href
  {https://doi.org/https://doi.org/10.1006/jsvi.1996.0471} {\bibfield
  {journal} {\bibinfo  {journal} {Journal of Sound and Vibration}\ }\textbf
  {\bibinfo {volume} {196}},\ \bibinfo {pages} {107} (\bibinfo {year}
  {1996})}\BibitemShut {NoStop}%
\bibitem [{\citenamefont {Thomsen}\ \emph {et~al.}(2019)\citenamefont
  {Thomsen}, \citenamefont {Moler\'on}, \citenamefont {Haag}, \citenamefont
  {van Manen},\ and\ \citenamefont {Robertsson}}]{Thomsen2019}%
  \BibitemOpen
  \bibfield  {author} {\bibinfo {author} {\bibfnamefont {H.~R.}\ \bibnamefont
  {Thomsen}}, \bibinfo {author} {\bibfnamefont {M.}~\bibnamefont {Moler\'on}},
  \bibinfo {author} {\bibfnamefont {T.}~\bibnamefont {Haag}}, \bibinfo {author}
  {\bibfnamefont {D.-J.}\ \bibnamefont {van Manen}},\ and\ \bibinfo {author}
  {\bibfnamefont {J.~O.~A.}\ \bibnamefont {Robertsson}},\ }\bibfield  {title}
  {\bibinfo {title} {Elastic immersive wave experimentation: Theory and
  physical implementation},\ }\href
  {https://doi.org/10.1103/PhysRevResearch.1.033203} {\bibfield  {journal}
  {\bibinfo  {journal} {Phys. Rev. Research}\ }\textbf {\bibinfo {volume}
  {1}},\ \bibinfo {pages} {033203} (\bibinfo {year} {2019})}\BibitemShut
  {NoStop}%
\bibitem [{\citenamefont {Vasmel}(2016)}]{Vasmel2016b}%
  \BibitemOpen
  \bibfield  {author} {\bibinfo {author} {\bibfnamefont {M.}~\bibnamefont
  {Vasmel}},\ }\emph {\bibinfo {title} {{Immersive boundary conditions for
  seismic wave propagation}}},\ \href
  {https://doi.org/10.3929/ethz-a-010677106} {Ph.D. thesis},\ \bibinfo
  {school} {ETH Z{\"{u}}rich, Switzerland} (\bibinfo {year} {2016})\BibitemShut
  {NoStop}%
\bibitem [{\citenamefont {B\"orsing}\ \emph {et~al.}(2019)\citenamefont
  {B\"orsing}, \citenamefont {Becker}, \citenamefont {Curtis}, \citenamefont
  {van Manen}, \citenamefont {Haag},\ and\ \citenamefont
  {Robertsson}}]{Boersing2019}%
  \BibitemOpen
  \bibfield  {author} {\bibinfo {author} {\bibfnamefont {N.}~\bibnamefont
  {B\"orsing}}, \bibinfo {author} {\bibfnamefont {T.~S.}\ \bibnamefont
  {Becker}}, \bibinfo {author} {\bibfnamefont {A.}~\bibnamefont {Curtis}},
  \bibinfo {author} {\bibfnamefont {D.-J.}\ \bibnamefont {van Manen}}, \bibinfo
  {author} {\bibfnamefont {T.}~\bibnamefont {Haag}},\ and\ \bibinfo {author}
  {\bibfnamefont {J.~O.}\ \bibnamefont {Robertsson}},\ }\bibfield  {title}
  {\bibinfo {title} {Cloaking and holography experiments using immersive
  boundary conditions},\ }\href
  {https://doi.org/10.1103/PhysRevApplied.12.024011} {\bibfield  {journal}
  {\bibinfo  {journal} {Phys. Rev. Applied}\ }\textbf {\bibinfo {volume}
  {12}},\ \bibinfo {pages} {024011} (\bibinfo {year} {2019})}\BibitemShut
  {NoStop}%
\bibitem [{\citenamefont {Becker}\ \emph {et~al.}(2020)\citenamefont {Becker},
  \citenamefont {B\"orsing}, \citenamefont {Haag}, \citenamefont {B\"arlocher},
  \citenamefont {Donahue}, \citenamefont {Curtis}, \citenamefont {Robertsson},\
  and\ \citenamefont {van Manen}}]{Becker2020}%
  \BibitemOpen
  \bibfield  {author} {\bibinfo {author} {\bibfnamefont {T.~S.}\ \bibnamefont
  {Becker}}, \bibinfo {author} {\bibfnamefont {N.}~\bibnamefont {B\"orsing}},
  \bibinfo {author} {\bibfnamefont {T.}~\bibnamefont {Haag}}, \bibinfo {author}
  {\bibfnamefont {C.}~\bibnamefont {B\"arlocher}}, \bibinfo {author}
  {\bibfnamefont {C.~M.}\ \bibnamefont {Donahue}}, \bibinfo {author}
  {\bibfnamefont {A.}~\bibnamefont {Curtis}}, \bibinfo {author} {\bibfnamefont
  {J.~O.~A.}\ \bibnamefont {Robertsson}},\ and\ \bibinfo {author}
  {\bibfnamefont {D.-J.}\ \bibnamefont {van Manen}},\ }\bibfield  {title}
  {\bibinfo {title} {Real-time immersion of physical experiments in virtual
  wave-physics domains},\ }\href
  {https://doi.org/10.1103/PhysRevApplied.13.064061} {\bibfield  {journal}
  {\bibinfo  {journal} {Phys. Rev. Applied}\ }\textbf {\bibinfo {volume}
  {13}},\ \bibinfo {pages} {064061} (\bibinfo {year} {2020})}\BibitemShut
  {NoStop}%
\bibitem [{\citenamefont {van Manen}\ \emph {et~al.}(2019)\citenamefont {van
  Manen}, \citenamefont {Moleron}, \citenamefont {Thomsen}, \citenamefont
  {Börsing}, \citenamefont {Becker}, \citenamefont {Haberman},\ and\
  \citenamefont {Robertsson}}]{vanManen2019}%
  \BibitemOpen
  \bibfield  {author} {\bibinfo {author} {\bibfnamefont {D.-J.}\ \bibnamefont
  {van Manen}}, \bibinfo {author} {\bibfnamefont {M.}~\bibnamefont {Moleron}},
  \bibinfo {author} {\bibfnamefont {H.~R.}\ \bibnamefont {Thomsen}}, \bibinfo
  {author} {\bibfnamefont {N.}~\bibnamefont {Börsing}}, \bibinfo {author}
  {\bibfnamefont {T.~S.}\ \bibnamefont {Becker}}, \bibinfo {author}
  {\bibfnamefont {M.~R.}\ \bibnamefont {Haberman}},\ and\ \bibinfo {author}
  {\bibfnamefont {J.~O.}\ \bibnamefont {Robertsson}},\ }\bibfield  {title}
  {\bibinfo {title} {Immersive boundary conditions for meta-material
  experimentation},\ }\href {https://doi.org/10.1121/1.5136649} {\bibfield
  {journal} {\bibinfo  {journal} {The Journal of the Acoustical Society of
  America}\ }\textbf {\bibinfo {volume} {146}},\ \bibinfo {pages} {2786}
  (\bibinfo {year} {2019})},\ \Eprint
  {https://arxiv.org/abs/https://doi.org/10.1121/1.5136649}
  {https://doi.org/10.1121/1.5136649} \BibitemShut {NoStop}%
\bibitem [{\citenamefont {Thomsen}\ \emph {et~al.}(2021)\citenamefont
  {Thomsen}, \citenamefont {Koene}, \citenamefont {Robertsson},\ and\
  \citenamefont {van Manen}}]{Thomsen2021}%
  \BibitemOpen
  \bibfield  {author} {\bibinfo {author} {\bibfnamefont {H.~R.}\ \bibnamefont
  {Thomsen}}, \bibinfo {author} {\bibfnamefont {E.~F.~M.}\ \bibnamefont
  {Koene}}, \bibinfo {author} {\bibfnamefont {J.~O.~A.}\ \bibnamefont
  {Robertsson}},\ and\ \bibinfo {author} {\bibfnamefont {D.-J.}\ \bibnamefont
  {van Manen}},\ }\bibfield  {title} {\bibinfo {title} {{FD-injection-based
  elastic wavefield separation for open and closed configurations}},\ }\href
  {https://doi.org/10.1093/gji/ggab275} {\bibfield  {journal} {\bibinfo
  {journal} {Geophysical Journal International}\ }\textbf {\bibinfo {volume}
  {227}},\ \bibinfo {pages} {1646} (\bibinfo {year} {2021})},\ \Eprint
  {https://arxiv.org/abs/https://academic.oup.com/gji/article-pdf/227/3/1646/40085128/ggab275.pdf}
  {https://academic.oup.com/gji/article-pdf/227/3/1646/40085128/ggab275.pdf}
  \BibitemShut {NoStop}%
\bibitem [{\citenamefont {S{\o}nneland}\ \emph {et~al.}(1986)\citenamefont
  {S{\o}nneland}, \citenamefont {Berg}, \citenamefont {Eidsvig}, \citenamefont
  {Haugen}, \citenamefont {Fotland},\ and\ \citenamefont
  {Vestby}}]{Sonneland2005}%
  \BibitemOpen
  \bibfield  {author} {\bibinfo {author} {\bibfnamefont {L.}~\bibnamefont
  {S{\o}nneland}}, \bibinfo {author} {\bibfnamefont {L.~E.}\ \bibnamefont
  {Berg}}, \bibinfo {author} {\bibfnamefont {P.}~\bibnamefont {Eidsvig}},
  \bibinfo {author} {\bibfnamefont {A.}~\bibnamefont {Haugen}}, \bibinfo
  {author} {\bibfnamefont {B.}~\bibnamefont {Fotland}},\ and\ \bibinfo {author}
  {\bibfnamefont {J.}~\bibnamefont {Vestby}},\ }\bibfield  {title} {\bibinfo
  {title} {{2-D deghosting using vertical receiver arrays}},\ }in\ \href
  {https://doi.org/10.1190/1.1893019} {\emph {\bibinfo {booktitle} {SEG
  Technical Program Expanded Abstracts 1986}}},\ Vol.~\bibinfo {volume} {4}\
  (\bibinfo  {publisher} {Society of Exploration Geophysicists},\ \bibinfo
  {year} {1986})\ pp.\ \bibinfo {pages} {516--519}\BibitemShut {NoStop}%
\bibitem [{\citenamefont {Hardage}(1985)}]{Hardage1985}%
  \BibitemOpen
  \bibfield  {author} {\bibinfo {author} {\bibfnamefont {B.~A.}\ \bibnamefont
  {Hardage}},\ }\bibfield  {title} {\bibinfo {title} {Vertical seismic
  profiling},\ }\href {https://doi.org/10.1190/1.1487141} {\bibfield  {journal}
  {\bibinfo  {journal} {The Leading Edge}\ }\textbf {\bibinfo {volume} {4}},\
  \bibinfo {pages} {59} (\bibinfo {year} {1985})},\ \Eprint
  {https://arxiv.org/abs/https://doi.org/10.1190/1.1487141}
  {https://doi.org/10.1190/1.1487141} \BibitemShut {NoStop}%
\bibitem [{\citenamefont {{PI Ceramic GmbH}}(2018)}]{Pica}%
  \BibitemOpen
  \bibfield  {author} {\bibinfo {author} {\bibnamefont {{PI Ceramic GmbH}}},\
  }\href {https://www.piceramic.com/en/} {\bibinfo {title} {{PICA Shear
  Actuators}}} (\bibinfo {year} {2018})\BibitemShut {NoStop}%
\bibitem [{\citenamefont {{Polytec GmbH}}(2019)}]{Polytec}%
  \BibitemOpen
  \bibfield  {author} {\bibinfo {author} {\bibnamefont {{Polytec GmbH}}},\
  }\href {https://www.polytec.com} {\bibinfo {title} {{PSV-500-3D Scanning
  Vibrometer}}} (\bibinfo {year} {2019})\BibitemShut {NoStop}%
\bibitem [{\citenamefont {Tsakmakidis}\ \emph {et~al.}(2007)\citenamefont
  {Tsakmakidis}, \citenamefont {Boardman},\ and\ \citenamefont
  {Hess}}]{Tsakmakidis2007}%
  \BibitemOpen
  \bibfield  {author} {\bibinfo {author} {\bibfnamefont {K.}~\bibnamefont
  {Tsakmakidis}}, \bibinfo {author} {\bibfnamefont {A.}~\bibnamefont
  {Boardman}},\ and\ \bibinfo {author} {\bibfnamefont {O.}~\bibnamefont
  {Hess}},\ }\bibfield  {title} {\bibinfo {title} {Nature: Trapped rainbow
  storage of light in metamaterials},\ }\href
  {https://doi.org/10.1038/nature06285} {\bibfield  {journal} {\bibinfo
  {journal} {Nature}\ }\textbf {\bibinfo {volume} {450}},\ \bibinfo {pages}
  {397} (\bibinfo {year} {2007})}\BibitemShut {NoStop}%
\bibitem [{\citenamefont {De~Ponti}\ \emph
  {et~al.}(2021{\natexlab{b}})\citenamefont {De~Ponti}, \citenamefont {Iorio},
  \citenamefont {Riva}, \citenamefont {Ardito}, \citenamefont {Braghin},\ and\
  \citenamefont {Corigliano}}]{DePonti2021}%
  \BibitemOpen
  \bibfield  {author} {\bibinfo {author} {\bibfnamefont {J.~M.}\ \bibnamefont
  {De~Ponti}}, \bibinfo {author} {\bibfnamefont {L.}~\bibnamefont {Iorio}},
  \bibinfo {author} {\bibfnamefont {E.}~\bibnamefont {Riva}}, \bibinfo {author}
  {\bibfnamefont {R.}~\bibnamefont {Ardito}}, \bibinfo {author} {\bibfnamefont
  {F.}~\bibnamefont {Braghin}},\ and\ \bibinfo {author} {\bibfnamefont
  {A.}~\bibnamefont {Corigliano}},\ }\bibfield  {title} {\bibinfo {title}
  {Selective mode conversion and rainbow trapping via graded elastic
  waveguides},\ }\href {https://doi.org/10.1103/PhysRevApplied.16.034028}
  {\bibfield  {journal} {\bibinfo  {journal} {Phys. Rev. Applied}\ }\textbf
  {\bibinfo {volume} {16}},\ \bibinfo {pages} {034028} (\bibinfo {year}
  {2021}{\natexlab{b}})}\BibitemShut {NoStop}%
\bibitem [{\citenamefont {Van~Damme}\ and\ \citenamefont
  {Zemp}(2018)}]{van2018measuring}%
  \BibitemOpen
  \bibfield  {author} {\bibinfo {author} {\bibfnamefont {B.}~\bibnamefont
  {Van~Damme}}\ and\ \bibinfo {author} {\bibfnamefont {A.}~\bibnamefont
  {Zemp}},\ }\bibfield  {title} {\bibinfo {title} {Measuring dispersion curves
  for bending waves in beams: a comparison of spatial fourier transform and
  inhomogeneous wave correlation},\ }\href@noop {} {\bibfield  {journal}
  {\bibinfo  {journal} {Acta Acustica united with Acustica}\ }\textbf {\bibinfo
  {volume} {104}},\ \bibinfo {pages} {228} (\bibinfo {year}
  {2018})}\BibitemShut {NoStop}%
\bibitem [{\citenamefont {{COMSOL AB}}(2022)}]{COMSOL}%
  \BibitemOpen
  \bibfield  {author} {\bibinfo {author} {\bibnamefont {{COMSOL AB}}},\ }\href
  {www.comsol.com} {\bibinfo {title} {{COMSOL Multiphysics v. 5.6}}} (\bibinfo
  {year} {2022})\BibitemShut {NoStop}%
\bibitem [{\citenamefont {Lemoult}\ \emph {et~al.}(2010)\citenamefont
  {Lemoult}, \citenamefont {Lerosey}, \citenamefont {de~Rosny},\ and\
  \citenamefont {Fink}}]{Lemoult2010}%
  \BibitemOpen
  \bibfield  {author} {\bibinfo {author} {\bibfnamefont {F.}~\bibnamefont
  {Lemoult}}, \bibinfo {author} {\bibfnamefont {G.}~\bibnamefont {Lerosey}},
  \bibinfo {author} {\bibfnamefont {J.}~\bibnamefont {de~Rosny}},\ and\
  \bibinfo {author} {\bibfnamefont {M.}~\bibnamefont {Fink}},\ }\bibfield
  {title} {\bibinfo {title} {Resonant metalenses for breaking the diffraction
  barrier},\ }\href {https://doi.org/10.1103/PhysRevLett.104.203901} {\bibfield
   {journal} {\bibinfo  {journal} {Phys. Rev. Lett.}\ }\textbf {\bibinfo
  {volume} {104}},\ \bibinfo {pages} {203901} (\bibinfo {year}
  {2010})}\BibitemShut {NoStop}%
\bibitem [{\citenamefont {Lemoult}\ \emph {et~al.}(2011)\citenamefont
  {Lemoult}, \citenamefont {Fink},\ and\ \citenamefont
  {Lerosey}}]{Lemoult2011}%
  \BibitemOpen
  \bibfield  {author} {\bibinfo {author} {\bibfnamefont {F.}~\bibnamefont
  {Lemoult}}, \bibinfo {author} {\bibfnamefont {M.}~\bibnamefont {Fink}},\ and\
  \bibinfo {author} {\bibfnamefont {G.}~\bibnamefont {Lerosey}},\ }\bibfield
  {title} {\bibinfo {title} {Acoustic resonators for far-field control of sound
  on a subwavelength scale},\ }\href
  {https://doi.org/10.1103/PhysRevLett.107.064301} {\bibfield  {journal}
  {\bibinfo  {journal} {Phys. Rev. Lett.}\ }\textbf {\bibinfo {volume} {107}},\
  \bibinfo {pages} {064301} (\bibinfo {year} {2011})}\BibitemShut {NoStop}%
\bibitem [{\citenamefont {Vered}\ and\ \citenamefont
  {Bucher}(2021)}]{Vered2021}%
  \BibitemOpen
  \bibfield  {author} {\bibinfo {author} {\bibfnamefont {Y.}~\bibnamefont
  {Vered}}\ and\ \bibinfo {author} {\bibfnamefont {I.}~\bibnamefont {Bucher}},\
  }\bibfield  {title} {\bibinfo {title} {Experimental multimode traveling waves
  identification in an acoustic waveguide},\ }\href
  {https://doi.org/https://doi.org/10.1016/j.ymssp.2020.107515} {\bibfield
  {journal} {\bibinfo  {journal} {Mechanical Systems and Signal Processing}\
  }\textbf {\bibinfo {volume} {153}},\ \bibinfo {pages} {107515} (\bibinfo
  {year} {2021})}\BibitemShut {NoStop}%
\bibitem [{\citenamefont {Becker}\ \emph {et~al.}(2021)\citenamefont {Becker},
  \citenamefont {van Manen}, \citenamefont {Haag}, \citenamefont {Bärlocher},
  \citenamefont {Li}, \citenamefont {Börsing}, \citenamefont {Curtis},
  \citenamefont {Serra-Garcia},\ and\ \citenamefont {Robertsson}}]{Becker2021}%
  \BibitemOpen
  \bibfield  {author} {\bibinfo {author} {\bibfnamefont {T.~S.}\ \bibnamefont
  {Becker}}, \bibinfo {author} {\bibfnamefont {D.-J.}\ \bibnamefont {van
  Manen}}, \bibinfo {author} {\bibfnamefont {T.}~\bibnamefont {Haag}}, \bibinfo
  {author} {\bibfnamefont {C.}~\bibnamefont {Bärlocher}}, \bibinfo {author}
  {\bibfnamefont {X.}~\bibnamefont {Li}}, \bibinfo {author} {\bibfnamefont
  {N.}~\bibnamefont {Börsing}}, \bibinfo {author} {\bibfnamefont
  {A.}~\bibnamefont {Curtis}}, \bibinfo {author} {\bibfnamefont
  {M.}~\bibnamefont {Serra-Garcia}},\ and\ \bibinfo {author} {\bibfnamefont
  {J.~O.~A.}\ \bibnamefont {Robertsson}},\ }\bibfield  {title} {\bibinfo
  {title} {Broadband acoustic invisibility and illusions},\ }\href
  {https://doi.org/10.1126/sciadv.abi9627} {\bibfield  {journal} {\bibinfo
  {journal} {Science Advances}\ }\textbf {\bibinfo {volume} {7}},\ \bibinfo
  {pages} {eabi9627} (\bibinfo {year} {2021})},\ \Eprint
  {https://arxiv.org/abs/https://www.science.org/doi/pdf/10.1126/sciadv.abi9627}
  {https://www.science.org/doi/pdf/10.1126/sciadv.abi9627} \BibitemShut
  {NoStop}%
\bibitem [{\citenamefont {Abma}\ \emph {et~al.}(2005)\citenamefont {Abma},
  \citenamefont {Kabir}, \citenamefont {Matson}, \citenamefont {Michell},
  \citenamefont {Shaw},\ and\ \citenamefont {McLain}}]{Abma2005}%
  \BibitemOpen
  \bibfield  {author} {\bibinfo {author} {\bibfnamefont {R.}~\bibnamefont
  {Abma}}, \bibinfo {author} {\bibfnamefont {N.}~\bibnamefont {Kabir}},
  \bibinfo {author} {\bibfnamefont {K.~H.}\ \bibnamefont {Matson}}, \bibinfo
  {author} {\bibfnamefont {S.}~\bibnamefont {Michell}}, \bibinfo {author}
  {\bibfnamefont {S.~A.}\ \bibnamefont {Shaw}},\ and\ \bibinfo {author}
  {\bibfnamefont {B.}~\bibnamefont {McLain}},\ }\bibfield  {title} {\bibinfo
  {title} {Comparisons of adaptive subtraction methods for multiple
  attenuation},\ }\href {https://doi.org/10.1190/1.1895312} {\bibfield
  {journal} {\bibinfo  {journal} {The Leading Edge}\ }\textbf {\bibinfo
  {volume} {24}},\ \bibinfo {pages} {277} (\bibinfo {year} {2005})},\ \Eprint
  {https://arxiv.org/abs/https://doi.org/10.1190/1.1895312}
  {https://doi.org/10.1190/1.1895312} \BibitemShut {NoStop}%
\bibitem [{\citenamefont {Li}\ \emph {et~al.}(2022)\citenamefont {Li},
  \citenamefont {Robertsson}, \citenamefont {Curtis},\ and\ \citenamefont {van
  Manen}}]{Li2022}%
  \BibitemOpen
  \bibfield  {author} {\bibinfo {author} {\bibfnamefont {X.}~\bibnamefont
  {Li}}, \bibinfo {author} {\bibfnamefont {J.}~\bibnamefont {Robertsson}},
  \bibinfo {author} {\bibfnamefont {A.}~\bibnamefont {Curtis}},\ and\ \bibinfo
  {author} {\bibfnamefont {D.-J.}\ \bibnamefont {van Manen}},\ }\bibfield
  {title} {\bibinfo {title} {Internal absorbing boundary conditions for
  closed-aperture wavefield decomposition in solid media with unknown
  interiors},\ }\href {https://doi.org/10.1121/10.0012578} {\bibfield
  {journal} {\bibinfo  {journal} {The Journal of the Acoustical Society of
  America}\ }\textbf {\bibinfo {volume} {152}},\ \bibinfo {pages} {313}
  (\bibinfo {year} {2022})},\ \Eprint
  {https://arxiv.org/abs/https://doi.org/10.1121/10.0012578}
  {https://doi.org/10.1121/10.0012578} \BibitemShut {NoStop}%
\bibitem [{\citenamefont {Pu}\ \emph {et~al.}(2022)\citenamefont {Pu},
  \citenamefont {Palermo},\ and\ \citenamefont {Marzani}}]{Pu2022}%
  \BibitemOpen
  \bibfield  {author} {\bibinfo {author} {\bibfnamefont {X.}~\bibnamefont
  {Pu}}, \bibinfo {author} {\bibfnamefont {A.}~\bibnamefont {Palermo}},\ and\
  \bibinfo {author} {\bibfnamefont {A.}~\bibnamefont {Marzani}},\ }\bibfield
  {title} {\bibinfo {title} {Topological edge states of quasiperiodic elastic
  metasurfaces},\ }\href
  {https://doi.org/https://doi.org/10.1016/j.ymssp.2022.109478} {\bibfield
  {journal} {\bibinfo  {journal} {Mechanical Systems and Signal Processing}\
  }\textbf {\bibinfo {volume} {181}},\ \bibinfo {pages} {109478} (\bibinfo
  {year} {2022})}\BibitemShut {NoStop}%
\bibitem [{\citenamefont {Graff}(2012)}]{graff2012wave}%
  \BibitemOpen
  \bibfield  {author} {\bibinfo {author} {\bibfnamefont {K.~F.}\ \bibnamefont
  {Graff}},\ }\href@noop {} {\emph {\bibinfo {title} {Wave motion in elastic
  solids}}}\ (\bibinfo  {publisher} {Courier Corporation},\ \bibinfo {year}
  {2012})\BibitemShut {NoStop}%
\end{thebibliography}%

\end{document}